\documentclass[9pt,twocolumn,twoside]{pnas-new}
\templatetype{pnasresearcharticle}
\setboolean{displaywatermark}{false}

\fancypagestyle{firststyle}{
   \fancyfoot[R]{\thepage}
   \fancyfoot[L]{}
}
\usepackage{bm} 					
\usepackage{upgreek}				
\usepackage{dsfont}
\usepackage{graphicx} 				
	\usepackage{color}				
	\usepackage{transparent}		
\usepackage{fancyhdr}				

\usepackage{setspace}

\usepackage{dsfont}				
\usepackage{microtype}		

\let\SQRT\sqrt
\renewcommand{\sqrt}[1]{\ensuremath{\SQRT{#1} \;}}

\let\REALPART\Re
\renewcommand{\Re}[1]{\ensuremath{\REALPART\left\{ #1 \right\} }}
\let\IMAGPART\Im
\renewcommand{\Im}[1]{\ensuremath{\IMAGPART\left\{ #1 \right\} }}

\catcode`_=\active
\newcommand_[1]{\ensuremath{\sb{\mathrm{#1}}}}

\newcommand{\intfin}[3]{\ensuremath{\int\sb{#2}^{#3}} \textup{d}#1 \;} 

\renewcommand{\mod}[1]{\ensuremath{\left| #1 \right|}} 
\newcommand{\td}{\ensuremath{\left(t\right)}} 
\newcommand{\Tr}[1]{\ensuremath{\textup{Tr}\left\{ #1 \right\}}} 
\newcommand{\trans}[1]{\ensuremath{ \left.#1\right.^{\mathrm{T}}}} 
\newcommand{\diag}[1]{\ensuremath{ \mathrm{diag}\!\left\{ #1 \right\} }}
\newcommand{\identity}{\ensuremath{ \mathds{1} }} 

\newcommand{\Br}[1]{\ensuremath{\left( #1 \right)}} 
\newcommand{\Sq}[1]{\ensuremath{\left[ #1 \right]}} 
\newcommand{\Cu}[1]{\ensuremath{\left\{ #1 \right\}}} 
\newcommand{\ghost}[1]{\ensuremath{\left. #1 \right.}} 
\newcommand{\upBr}[2]{\ensuremath{\ghost{#1^{\Br{#2}}}} } 
\renewcommand{\mod}[1]{\ensuremath{\left| #1 \right|}}	

\newcommand{\expect}[1]{\ensuremath{\left< #1 \right>}}
\newcommand{\comm}[2]{\ensuremath{\left[ #1,#2 \right]}}
\newcommand{\n}{\ensuremath{\bar{n}}} 

\newcommand{\I}{\mathrm{i}} 									
\newcommand{\e}{\mathrm{e}} 								
\newcommand{\RWA}{\ensuremath{\mathrm{RWA}}} 	
\newcommand{\COP}{\ensuremath{\textup{COP}}} 	

\newcommand{\cO}{{\mathcal{O} }} 

\newcommand{\half}{\ensuremath{\frac{1}{2}}}   

\newcommand{\FP}{Fabry-P\'{e}rot } 

\newcommand{\via}{\textit{via}} 
\newcommand{\ie}{\textit{i.e.}} 
\newcommand{\eg}{\textit{e.g.}} 
\newcommand{\etc}{\textit{etc.}} 
\newcommand{\vv}{\textit{vice versa}} 
\newcommand{\viz}{\textit{viz}.} 
\newcommand{\cf}{\textit{cf}.} 
\newcommand{\ala}{\textit{\`{a} la}} 
\renewcommand{\etal}{\textit{et al.}} 

\title{A Quantum Heat Machine from Fast Optomechanics}

\author[a,1]{James S. Bennett}
\author[a]{Lars S. Madsen}
\author[a]{Halina Rubinsztein-Dunlop}
\author[a]{Warwick P. Bowen}
\affil[a]{Australian Research Council Centre of Excellence for Engineered Quantum Systems (EQuS), School of Mathematics and Physics, The University of Queensland, St Lucia, QLD 4072, Australia}
\leadauthor{Bennett}

\authorcontributions{W.P.B. conceived initial concept. J.S.B. and L.S.M. performed calculations. W.P.B. provided oversight. H.R.D. provided general advice. All authors were involved in writing.}
\correspondingauthor{\textsuperscript{1}To whom correspondence should be addressed. E-mail: james.bennett2@uqconnect.edu.au}


\begin{abstract}
We consider a thermodynamic machine in which the working fluid is a quantized harmonic oscillator that is controlled on timescales that are much faster than the oscillator period. We find that operation in this `fast' regime allows access to a range of quantum thermodynamical behaviors that are otherwise inaccessible, including heat engine and refrigeration modes of operation, quantum squeezing, and transient cooling to temperatures below that of the cold bath. The machine involves rapid periodic squeezing operations and could potentially be constructed using pulsed optomechanical interactions. The prediction of rich behavior in the fast regime opens up new possibilities for quantum optomechanical machines and quantum thermodynamics.
\end{abstract}

\dates{This manuscript was compiled on \today}

\begin{document}

\maketitle
\thispagestyle{firststyle}
\ifthenelse{\boolean{shortarticle}}{\ifthenelse{\boolean{singlecolumn}}{\abscontentformatted}{\abscontent}}{}

\dropcap{T}he union of thermodynamics and quantum mechanics has proved extremely fruitful since its earliest days. Recent years have seen a rapid acceleration of this progress \cite{Millen2015,Vinjanampathy2017,Anders2017}, including important developments such as the generalization of the second law of thermodynamics to the quantum realm \cite{Brandao2015}, as well as the first general proof of the third law of thermodynamics \cite{Masanes2017}. Tools from quantum information theory \cite{Goold2016} have also clarified the role of information in thermodynamical processes. As one of the most significant applications of thermodynamics, heat machines---that is, engines, refrigerators, and heat pumps---have been a natural area of focus~\cite{Alicki2014,Alicki2018}. Much like their classical cousins, these heat machines are playing an important role in elucidating the fundamental laws of thermodynamics. For instance, the quantum Szilard engine \cite{Kim2011,Cai2012,Park2013,Mohammady2017} has proven a useful playground for investigating the effects of quantum non-locality \cite{Kim2011,Cai2012}, indistinguishability \cite{Kim2011,Cai2012}, and measurement and feedback \cite{Mohammady2017} in thermodynamics; experiments are also beginning to enter the regime where these theoretical predictions can be sensitively tested \cite{Koski2014}.

In recent years, it has become possible to bring mechanical oscillators fully into the quantum regime \via{} strong, resonantly-enhanced interactions with light or microwaves  (\eg{} \cite{Chan2011a,Teufel2011,Aspelmeyer2014review}). As such, opto- and electromechanical systems are promising testbeds for quantum thermodynamic machines. Nanomechanical oscillators have already been used to construct classical thermal engines~\cite{Steeneken2011} and to perform proof-of-principle experiments investigating the effects of non-equilibrium squeezed thermal reservoirs on an engine's efficiency~\cite{Klaers2017}. In this paper we investigate a new regime for oscillator-based quantum thermal machines, where the dynamics of the machine are controlled on timescales that are fast compared to the period of the oscillator. In particular, we consider a viscously-damped oscillator subjected to a fast periodic train of impulsive squeezing operations. We show that this machine is capable of a range of thermodynamical behaviors, including heat engine, refrigerator, and heat pump modes of operation. The working fluid (oscillator) can be quantum squeezed during portions of the thermal cycle. The oscillator can also be transiently cooled to below the cold bath temperature, unlike standard techniques for cooling an oscillator~\cite{Mancini1998,Schliesser2009}. Interestingly, these behaviors all hinge on the sub-mechanical-period (`fast') dynamics of viscous damping. While applicable to oscillator-based quantum machines quite generally, our machine might---for example---be experimentally implemented using a quantum electromechanical oscillator coupled to a pulsed electromagnetic field. Within this context, we provide one possible protocol to implement the machine \via{} a sequence of pulsed optomechanical interactions~\cite{Vanner2011, Bennett2016, Khosla2017, Bennett2018}. Our analysis suggests that this protocol is feasible with challenging but plausible extensions of the current state-of-the-art.

The existence of previously unanticipated, rich thermodynamical behaviors in the regime of fast dynamics, shown here for harmonic oscillators (\cf{} \cite{Mukherjee2020}), opens up a new and potentially fruitful area of exploration for quantum thermodynamics, particularly with regards to finite-time, non-equilibrium machines. 

\section*{Quantum Heat Machines} \label{Sec:HeatMachines}

Broadly, \textit{quantum} heat machines can be classified as those that are either \textit{a}) built from intrinsically quantized systems, such as spins or atoms; or \textit{b}) those for which quantum mechanics is required for a complete description of their operation. Numerous designs for quantum heat machines can be found in the literature as far back as 1967, when the three-level maser was first modeled as such \cite{Geusic1967}. Since then, many systems have been proposed for use as quantum thermal machines, \eg{} two-level systems \cite{Geva1992,GelbwaserKlimovsky2013,Mukherjee2020}, multi-level systems \cite{Geusic1967,Quan2005}, and harmonic oscillators \cite{JianHui2007,delCampo2014,Rezek2006,Hofer2016,Insinga2016} (amongst others, \eg{} \cite{Quan2007,Quan2009}). These have been used to investigate everything from fundamental questions---such as: what is the smallest engine that can be built~\cite{Kim2011,GelbwaserKlimovsky2013,Rossnagel2016}?; what properties should a quantum heat bath have \cite{Ford1988}?; how does thermalization occur \cite{LeCunuder2016,Ronzani2018}?; and how do quantum correlations alter or improve an engine's performance~\cite{Arnaud2002,Scully2003,Ghosh2017}?---to practical demonstrations, driven by advances in quantum engineering that allow dynamic control of an increasing range of quantum systems at the required level, \eg{}~collective atomic spin oscillators \cite{Kohler2017}, trapped ions \cite{Maslennikov2019}, semiconductor quantum dots \cite{Josefsson2018}, superconducting circuits \cite{Chen2012,Manikandan2019}, and nitrogen--vacancy centers in diamond \cite{Klatzow2019}.

Two broad classes of machines have been discussed in the optomechanics literature: `traditional' machines that operate by performing multiple strokes---as we are accustomed to in classical thermodynamics---and those that operate autonomously (\ie{} with no external, classical controller). Optomechanical polariton heat machines \cite{Zhang2014, Dong2015}, Otto engines \cite{delCampo2014}, and certain levitated optomechanical engines \cite{Dechant2015} fall into the former category. In the latter category are engines that leverage non-classical initial states as a resource \cite{Gelbwaser-Klimovsky2014}, and those that are continuously coupled to two optical baths at different temperatures \cite{Mari2015,Roulet2017}. Other studies have explored isolated operations that could be incorporated into an optomechanical thermodynamic cycle, such as calculating the work performed by an isolated harmonic oscillator subject to a frequency shift (either step-like or periodic) \cite{Deffner2008,Galve2009}, or the work generated by a sudden quench of the optomechanical interaction strength \cite{Brunelli2015}.

Our proposal is a form of multi-stroke optomechanical heat machine. It is distinguished from previous proposals in three key ways. Most significantly,  the machine operates in the `fast' regime, where each thermodynamic cycle is achieved in a time much shorter than the mechanical period; this permits otherwise unachievable behaviors. Secondly, work is injected and extracted using squeezing operations \cite{Bennett2018}, rather than displacements or symmetrical manipulations of the mechanical noise (cooling and heating). This means that the mechanical state is always non-equilibrium, and can in principle be squeezed to below the ground state variance (\ie{} made non-classical \cite{Dodonov2002}). The squeezing is achieved without any change to the mechanical frequency, unlike other proposals \cite{Dechant2015,Deffner2008,Galve2009,Ian2014}; as a result, the bath occupancies are unambiguous and fixed throughout the cycle. Finally, our proposal does not rely on non-equilibrium reservoirs, be they squeezed thermal reservoirs \cite{Klaers2017,Rossnagel2014, Huang2012,Niedenzu2018} or squeezed work reservoirs\footnote{Our suggested optomechanical implementation does make use of squeezed optical ancillae, but these are an artefact of the pulsed squeezer used, rather than a requirement of the underlying thermal cycle.} (`batteries') \cite{Ghosh2017}. Instead, the squeezing is applied directly to the working fluid.

\section*{Thermodynamic cycle} \label{Sec:Protocol}

\begin{figure}
	\centering
	\def\svgwidth{1\columnwidth}
	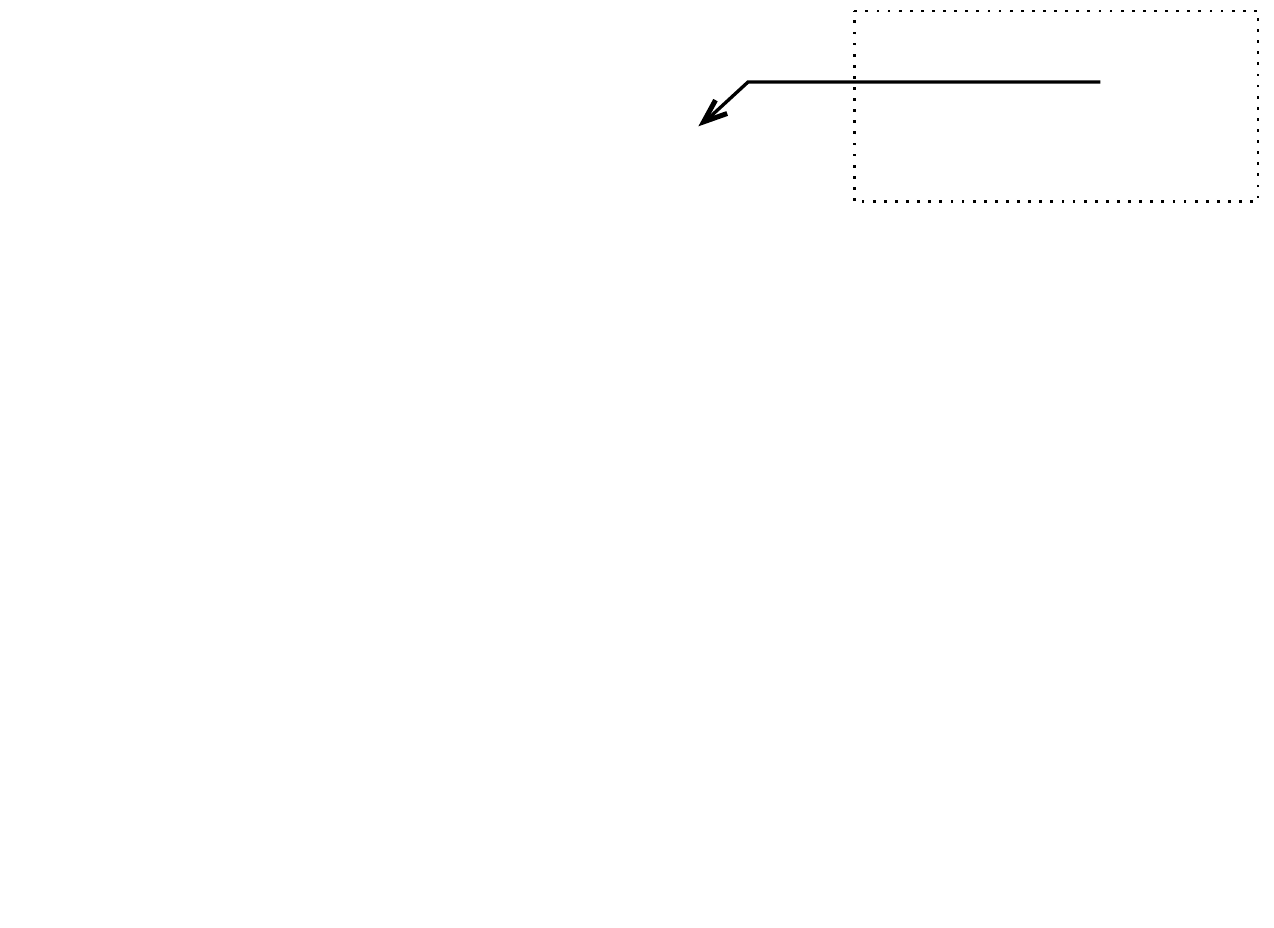
	\caption{\label{Fig:SequenceSimplified} Physical and temporal layout of the squeeze--rotate--unsqueeze thermal machine.\\
	\textit{a}) System schematic. A harmonic oscillator (centre) in state $\hat{\rho}$ is coupled to a hot bath ($\n_{H}$, orange) at a rate $\Gamma$. A work reservoir ($\mu$, dark grey) performs squeezing operations on the oscillator. Imperfections in the squeezer couples the oscillator to a cold bath ($\n_{C}$, teal).\\
	\textit{b}) Wigner representation of the thermal noise added during interaction with the hot bath ($V_{H}$). The black ellipse shows an equiprobability contour.\\
	\textit{c}) Model of imperfect squeezing operation $S\sb{j}^{\prime}$ as a perfect squeezer followed by an interaction with a cold bath. Both are modeled as being instantaneous.\\
	\textit{d}) Diagram of a single step of the squeeze--rotate--unsqueeze protocol. Beginning at the cyclical steady-state ($V_{CSS}$, left), the cycle proceeds clockwise. After execution of the first squeezer ($S_{1}^{\prime}$) the system is in state $\upBr{V}{2}$, then a time $\tau$ of interaction with the hot reservoir yields the state $\upBr{V}{3}$. Finally, the cycle is completed by the second squeezer ($S_{2}^{\prime}$), returning the oscillator to its cyclical steady-state. Illustrative Wigner functions of the aforementioned states ($V_{CSS}$, $\upBr{V}{2}$, and $\upBr{V}{3}$) are shown in purple.}
\end{figure}

Each cycle of our proposed protocol is divided into three broad steps as delineated in Fig.~\ref{Fig:SequenceSimplified}, panels \textit{a}) and \textit{d}). Firstly, an initial imperfect squeezing operation, $S_{1}^{\prime}$; then a short time $\tau$ of damped evolution whilst in contact with a hot bath; then a final imperfect squeezer, $S_{2}^{\prime}$. Imperfections in the squeezers introduce an effective cold bath. This `squeeze--rotate--unsqueeze' sequence is repeated at an angular rate of $\omega_{ap} = 2\pi/\tau$. To best demonstrate the operating principle of our `squeeze--rotate--unsqueeze' machine we will be considering the regime in which the squeezers are effectively instantaneous and the entire protocol repeats much more rapidly than the mechanical cycle, \ie{} $\omega_{ap} \gg \omega_{M}$, with $\omega_{M}$ being the mechanical (angular) frequency.

We select the operations during each cycle such that in the absence of decoherence the protocol simply reproduces free mechanical evolution (\ie{} phase space rotations). To achieve this, we set the first squeezer ($S_{1}$) such that it performs the operation
\begin{eqnarray}
	X & \rightarrow & \mu^{-1} X, \\
	P & \rightarrow & \mu P,
\end{eqnarray}
where $\mu$ is the squeezing strength, $X$ is the mechanical position normalised by the zero-point fluctuation length, and $P$ is the corresponding normalised momentum operator ($\comm{X}{P}=2\I$). In this notation, the momentum becomes antisqueezed for $\mu > 1$ and squeezed for $\mu < 1$. We choose the second squeezer to be $S_{2} = R S_{1}^{-1} R^{\mathrm{T}}$, where $R$ is a rotation matrix with angle $\omega_{M} \tau$. In the absence of loss, this yields the net effect $S_{2}RS_{1} = R$. We can therefore conclude that all non-trivial behaviors of our machine must be intrinsically linked to decoherence.

\subsection*{Hot bath coupling} \label{Sec:HotBath}

As already noted above, the non-trivial behaviors of our proposed machine are all tied directly to the decoherence processes governing the hot and cold thermal baths. We will consider two models of the interaction between a harmonic oscillator and thermal bath in this paper: viscous damping, and the Born-Markov master equation (BMME). In the former case loss is entirely dependent upon the mechanical momentum, whilst the latter shares losses equally between both $X$ and $P$ \cite{BowenMilburn}. Though these are by no means the only loss models that are physically relevant to harmonic oscillators \cite{Adhikari2001a,Adhikari2001b}, they accurately describe optomechanical experiments (in their respective realms of validity), and are the most commonly used in the optomechanical literature\footnote{See \eg{} the review by Aspelmeyer, Kippenberg, and Marquardt \cite{Aspelmeyer2014review}, which shows extensive use of both models.}. Importantly, the two different loss treatments yield dramatically different predictions, as we will demonstrate below.

\subsubsection*{Viscous damping} \label{Sec:HotBathViscous}

The Langevin equations describing an oscillator being viscously damped by a thermal bath are \cite{Aspelmeyer2014review}
\begin{eqnarray}
	\dot{X} & = & +\omega_{M} P, \label{Eqn:Xdot} \\
	\dot{P} & = & -\omega_{M} X - \Gamma P +\sqrt{2 \Gamma} \xi\td, \label{Eqn:Pdot}
\end{eqnarray}
where $\omega_{M}$ is the mechanical resonance frequency. The second term of \eqref{Eqn:Pdot} corresponds to viscous damping with an average energy decay rate of $\Gamma$, and the third describes the concomitant noise required by the fluctuation--dissipation theorem \cite{Vinjanampathy2017}. Note that these equations are asymmetric under rotations in phase space because the loss and thermal noise terms couple only to $P$.

\eqref{Eqn:Xdot} \& \eqref{Eqn:Pdot} can be derived by considering certain limits \cite{Giovannetti2001} of the independent oscillator model, the most general model of a linear, passive heat bath \cite{Ford1988}. There are two main requirements for them to be valid. Firstly, the bath temperature must be sufficiently large. This is because \eqref{Eqn:Xdot} and \eqref{Eqn:Pdot} do not correspond to a Lindblad form master equation, and so are not guaranteed to be completely positive for all initial states and bath temperatures \cite{Lindblad1976, Kohen1997}. For this reason, we will restrict ourselves to the high bath occupancy regime throughout this paper ($\n_{H}\geq 10^{2}$ in all instances). Secondly, Markovian equations of motion require weak coupling between the system and bath. The coupling strength is characterized by the $Q$ factor or `quality', $Q = \omega_{M}/\Gamma$; higher $Q$ corresponds to weaker coupling. $Q$ factors in excess of approximately ten are sufficient to permit use of Markovian equations of motion \cite{BowenMilburn,Giovannetti2001}; micro-- and nano-mechanical resonators have been reported to have $Q$ factors up to $\Br{9.8 \pm 0.2}\times 10^{7}$ \cite{Ghadimi2018,MacCabe2019}, with $Q > 10^{5}$ being readily achieved in many materials and resonator geometries \cite{Aspelmeyer2014review}. This justifies the use of Markovian equations of motion. This is in contrast to similar work \cite{Mukherjee2020} on rapidly-modulated two-level systems, where the bath cannot be modeled using Markovian dynamics.

\subsubsection*{Born-Markov approximation} \label{Sec:HotBathBMME}

A popular alternative to \eqref{Eqn:Xdot} and~\eqref{Eqn:Pdot} is the Born--Markov master equation (BMME), which corresponds to the Langevin equations \cite{Aspelmeyer2014review}
\begin{eqnarray}
	\dot{X} & = & +\omega_{M} P - \frac{\Gamma}{2} X + \sqrt{\Gamma} X_{in}\td, \label{Eqn:XdotRWA} \\
	\dot{P} & = & -\omega_{M} X - \frac{\Gamma}{2} P + \sqrt{\Gamma} P_{in}\td. \label{Eqn:PdotRWA}
\end{eqnarray}
In this case the loss and noise are shared equally between $X$ and $P$ (noting that the noise operators $X_{in}$ and $P_{in}$ have identical statistics).

\eqref{Eqn:XdotRWA} and \eqref{Eqn:PdotRWA}---much like the viscously damped equations---can be derived from the independent oscillator model. The BMME arises from making a rotating wave approximation (RWA) that discards all energy non-conserving terms from the system--bath Hamiltonian \cite{BowenMilburn}. Further technical notes on the relationship between \eqref{Eqn:Xdot} \& \eqref{Eqn:Pdot} and \eqref{Eqn:XdotRWA} \& \eqref{Eqn:PdotRWA} are available in Appendix \ref{Sec:RWAEOMs}.

\subsubsection*{Fast dynamics: the link with squeezing} \label{Sec:WhySqueeze}

Over long timescales, the viscously-damped and BMME models yield almost identical behaviors; notably, they converge to the same equilibrium state \cite{Kohen1997}. Conversely, they exhibit marked differences at short times. To see this, consider the oscillator's covariance matrix,
\begin{equation}
	V\td = \Br{\begin{array}{c c}
			\expect{\Br{X-\expect{X}}^{2}} & \Re{\expect{XP}}-\expect{X}\expect{P} \\
			\cdots & \expect{\Br{P-\expect{P}}^{2}}
			\end{array}},
			\label{Eqn:CovarDefinition}
\end{equation}
with $V = \trans{V}$ and $\Re{\expect{\cdots}}$ denoting the real part of the expectation value. For all times $t\geq 0$, $V\td$ is related to its initial value $V_{0}$ by a linear transformation.

For the momentum-damped equations of motion we obtain $V\td = M_{H}\td V_{0} M_{H}^{\mathrm{T}}\td + V_{H}\td$, where $M_{H}\td$ is a square matrix encoding the homogeneous part of the dynamics (\cf{} Appendix \ref{Sec:IndependentOscillator}). It may loosely be thought of as a lossy rotation matrix. The aggregated effect of thermal noise is described by the added noise covariance matrix $V_{H}\td$, which depends upon $M_{H}\td$ and the noise autocorrelation $\expect{\xi\td\xi\Br{t^{\prime}}+\xi\Br{t^{\prime}}\xi\td} = 2\Br{2\n_{H}+1}\delta\Br{t-t^{\prime}}$, where $\n_{H}$ is the equilibrium occupancy of the hot bath (\cf{}~Appendix~\ref{Sec:IndependentOscillator}). The occupancy is determined by the bath temperature $T_{H}$ through the Bose--Einstein distribution \cite{BowenMilburn}.

For short evolution times ($t \ll \omega_{M}^{-1}$) the noise added to the system by viscous damping into the thermal bath is
\begin{equation}
	V_{H}\Br{t \ll \omega_{M}^{-1}} = \Br{2\n_{H}+1} \Br{\begin{array}{c c}
							\frac{2}{3}\Gamma\omega_{M}^{2}t^{3} & \Gamma\omega_{M} t^{2} \\
							\Gamma \omega_{M} t^{2} & 2\Gamma t
							\end{array}}
			\label{Eqn:VFF}
\end{equation}
to leading non-trivial order in each matrix element. The first (second) diagonal element describes the noise added to $X$ ($P$). Remarkably, despite the fact that the bath is in a perfectly thermal state, the noise it introduces to the system is skewed heavily towards momentum noise over short interaction times (see Fig.~\ref{Fig:SequenceSimplified} \textit{b}) for a graphical representation). This is because the thermal noise only couples to momentum, not position.

A similar linear transformation holds for the RWA case, but with the replacements $M_{H}\rightarrow \upBr{M_{H}}{\mathrm{RWA}}$ and $V_{H} \rightarrow \upBr{V_{H}}{\mathrm{RWA}}$.  In stark contrast to \eqref{Eqn:VFF}, the noise added in the RWA is completely symmetrical between position and momentum; to leading order,
\[
	\upBr{V_{H}}{\mathrm{RWA}}\Br{t \ll \omega_{M}^{-1}} = \Br{2\n_{H}+1} \Gamma t \identity,
\]
with $\identity$ being the identity matrix. In fact, the noise remains symmetrical for all bath interaction times under the RWA. This is a result of the BMME coupling noise into both position and momentum.

\subsection*{Squeezer and cold bath coupling} \label{Sec:Squeezer}

As seen in Fig.~\ref{Fig:SequenceSimplified} \textit{c}), each imperfect squeezer  $S\sb{j}^{\prime} \; \Br{j = 1,2}$ can be thought of as being a combination of a unitary squeezer and a cold thermal bath (with occupancy $\n_{C} < \n_{H}$).  The unitary squeezer is $S\sb{j}$ (distinguished from $S\sb{j}^{\prime}$ by the lack of prime).

In order to clearly illustrate the essential physics underpinning our machine we have chosen to take the limit in which both of these processes are effectively instantaneous. Discussion of the practicalities of this limit is contained below (`Experimental Feasibility') and Appendix~\ref{App:NonRWABath}.

The interaction with the cold bath is modeled using \eqref{Eqn:Xdot} and~\eqref{Eqn:Pdot}, with an occupancy $\n_{C}$ (satisfying $1 \ll \n_{C} < \n_{H}$), decay rate $\gamma$, and interaction time $t_{C}$. To take the instantaneous interaction limit we fix the $\gamma t_{C}$ product, then let $t_{C} \rightarrow 0$. The resulting interaction is characterized by a new parameter, $\epsilon = 1-\e^{-\gamma t_{C}}$, such that $\epsilon = 0$ corresponds to no loss and $\epsilon = 1$ is maximally lossy. As with the hot bath, we represent the homogeneous part of the transformation as $M_{C}$, and the covariance of the added noise by $V_{C}$.
\begin{eqnarray}
	M_{C} & = & \Br{\begin{array}{cc} 1 & 0 \\ 0 & 1-\epsilon \end{array}}, \label{Eqn:MC} \\
	V_{C} & = & \Br{2\n_{C}+1}\Br{\begin{array}{cc} 0 & 0 \\ 0 & \epsilon\Br{2-\epsilon} \end{array}}. \label{Eqn:VC}
\end{eqnarray}
As expected from \eqref{Eqn:VFF}, the noise and loss only affect the momentum. Further details are given in Appendix~\ref{App:NonRWABath}.

If the BMME is used then $\upBr{M_{C}}{\RWA} = \sqrt{1-\epsilon}\identity$ and $\upBr{V_{C}}{\RWA} = \epsilon\Br{2\n_{C}+1}\identity$; note that both are symmetrical between $X$ and $P$, just as for the hot bath coupling.

Details of how such a squeezer could be realised will be discussed in `Experimental Feasibility', below.

\section*{Cyclical steady-state} \label{Sec:CyclicalState}

Having established the building blocks of our thermal cycle, let us turn to assembling them. A typical thermodynamic machine (of the non-autonomous variety) operates by performing a process in which the working fluid is returned to the same state at the end of each cycle. We will refer to this state, indicated in Fig.~\ref{Fig:SequenceSimplified}, panel~\textit{d}), as the `cyclical steady-state' (CSS). Repeated application of our squeezing protocol will gradually force any initial state towards a zero-mean, Gaussian CSS \footnote{This approach to steady-state is guaranteed because $\det\Cu{M_{hom}} < 1$.}. Therefore, the CSS is entirely characterized by its covariance matrix $V_{CSS}$, which is defined by the self-consistency condition
\begin{equation}
	V_{CSS} = M_{hom}\Br{\tau}V_{CSS}\trans{M}_{hom}\Br{\tau}  + V_{add}\Br{\tau},
	\label{Eqn:SteadyState}
\end{equation}
\ie{} the state returns to itself under a full cycle of the machine. The matrix $M_{hom}\Br{\tau} = M_{C}S_{2}M_{H}\Br{\tau} M_{C}S_{1}$ gives the homogeneous component of the evolution, and
\begin{eqnarray*}
	V_{add}\Br{\tau} & = & V_{C} +  M_{C}S_{2}V_{H}\Br{\tau}S_{2}^{\mathrm{T}}M_{C}^{\mathrm{T}} \\
	& & {} + M_{C}S_{2}M_{H}\Br{\tau}V_{C}M_{H}^{\mathrm{T}}\Br{\tau}S_{2}^{\mathrm{T}}M_{C}^{\mathrm{T}}
\end{eqnarray*}
is the aggregate effect of the noise entering from the baths over one cycle. Both matrices depend on the time over which a single cycle of the protocol is executed, $\tau = 2\pi/\omega_{ap}$. It is critical to note that the added noise is strongly modified by $S_{2}$. This allows us to manipulate the asymmetry of $V_{add}\Br{\tau}$ by adjusting the evolution time $\tau$ and squeezing strength $\mu$.

\eqref{Eqn:SteadyState} can be recognised as a Sylvester equation, which is readily solved using standard numerical techniques \cite{Aliev2017}.

\begin{figure}
	\centering
	\def\svgwidth{1\columnwidth}
	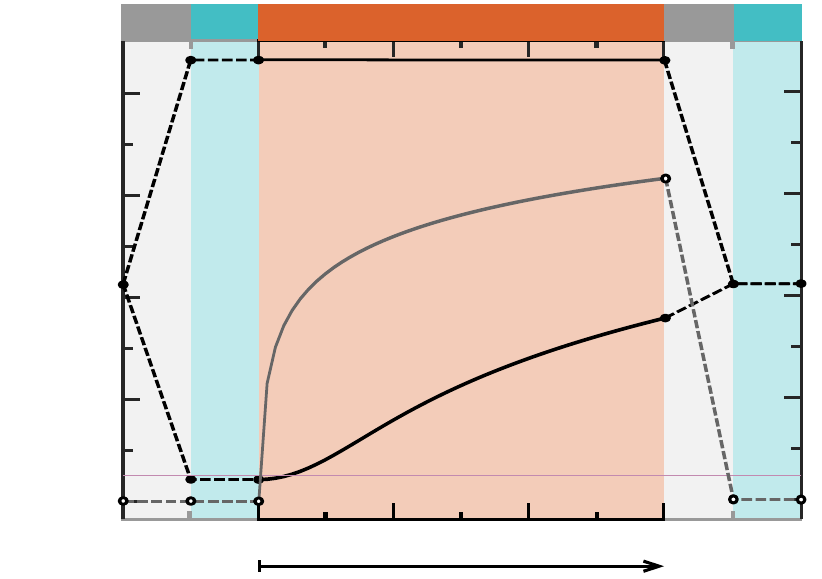
	\caption{\label{Fig:TimeDependence} Behavior of the covariance matrix over a single thermodynamic cycle of total time $\tau$. The system begins in an approximately symmetrical state, is position squeezed ($\mu = 9$, grey), and then interacts with the cold bath ($\epsilon = \pi\times 10^{-10}$, $\n_{C} = 180$, teal). It is then allowed to evolve whilst in contact with the hot bath ($Q=10^{4}$, $\n_{H} = 200$, $\omega_{ap}/\omega_{M} = 80$, orange); note the build-up of correlations ($V_{XP}$, dark grey line, unfilled markers) between position and momentum as the state evolves. Finally, the system is `unsqueezed' with $S_{2}^{\prime}$, returning it to the initial state. Crucially, note that the oscillator is quantum squeezed ($V_{XX}=0.88$) at points in the cycle. No time elapses along the dashed lines. The vertical scaling allows simultaneous display of the large variances and small covariance; $\mathrm{asinh}\Br{V/2}$ is essentially linear for small $V$, but becomes logarithmic for large $V$.}
\end{figure}

Armed with the CSS, we can then chart out the oscillator's evolution over the course of a single cycle of the protocol. Fig.~\ref{Fig:TimeDependence} shows an exemplary calculation. On the left, the oscillator begins in the CSS, which has symmetric variances ($V_{XX} = V_{PP}$ and zero covariance). The unitary squeezer $S_{1}$ then amplifies the momentum variance, correspondingly attenuating $V_{XX}$. Noise and loss are then introduced by the cold bath, though this effect is not visible to the eye for the parameters used in Fig.~\ref{Fig:TimeDependence} ($\epsilon = \pi\times 10^{-10}$, $\n_{C} = 180$). Note that the variance of the position is less than that of the ground state at this stage, \ie{} the state is quantum squeezed. The oscillator then undergoes a time of damped evolution whilst in contact with the hot bath ($Q=10^{4}$, $\n_{H} = 200$). As it does so, the momentum and position are exchanged, resulting in the build-up of correlations between them ($V_{XP}$ increases). At the conclusion of this step, the squeezer $S_{2}$ returns the system to a nearly symmetrical state. Finally, a small amount of noise enters from the cold bath to complete the cycle. As expected, the oscillator returns back to its CSS.

As already noted, the parameters used in Fig.~\ref{Fig:TimeDependence} were chosen to demonstrate an important feature of our thermal machine; the working fluid can be in a nonclassical (\ie{} quantum squeezed \cite{Dodonov2002}) state at the conclusion of $S_{1}^{\prime}$. 

\subsection*{Effective occupancy} \label{Sec:EffectiveOccupancy}

Numerical solutions of \eqref{Eqn:SteadyState} show that the CSS---the state at the point in the cycle indicated in Fig.~\ref{Fig:SequenceSimplified}, panel~\textit{d})---is very well approximated by a thermal-like state, $V_{CSS} = \Br{2\n_{CSS}+1}\identity$, where $\n_{CSS}$ is the effective CSS occupancy (\cf{} Appendix~\ref{App:SteadyState}). For example, the CSS shown in Fig.~\ref{Fig:TimeDependence} has this form, with $V_{XX} = V_{PP}$ and $V_{XP} \approx 0$.

\begin{figure}
	\centering
	\def\svgwidth{1\columnwidth}
	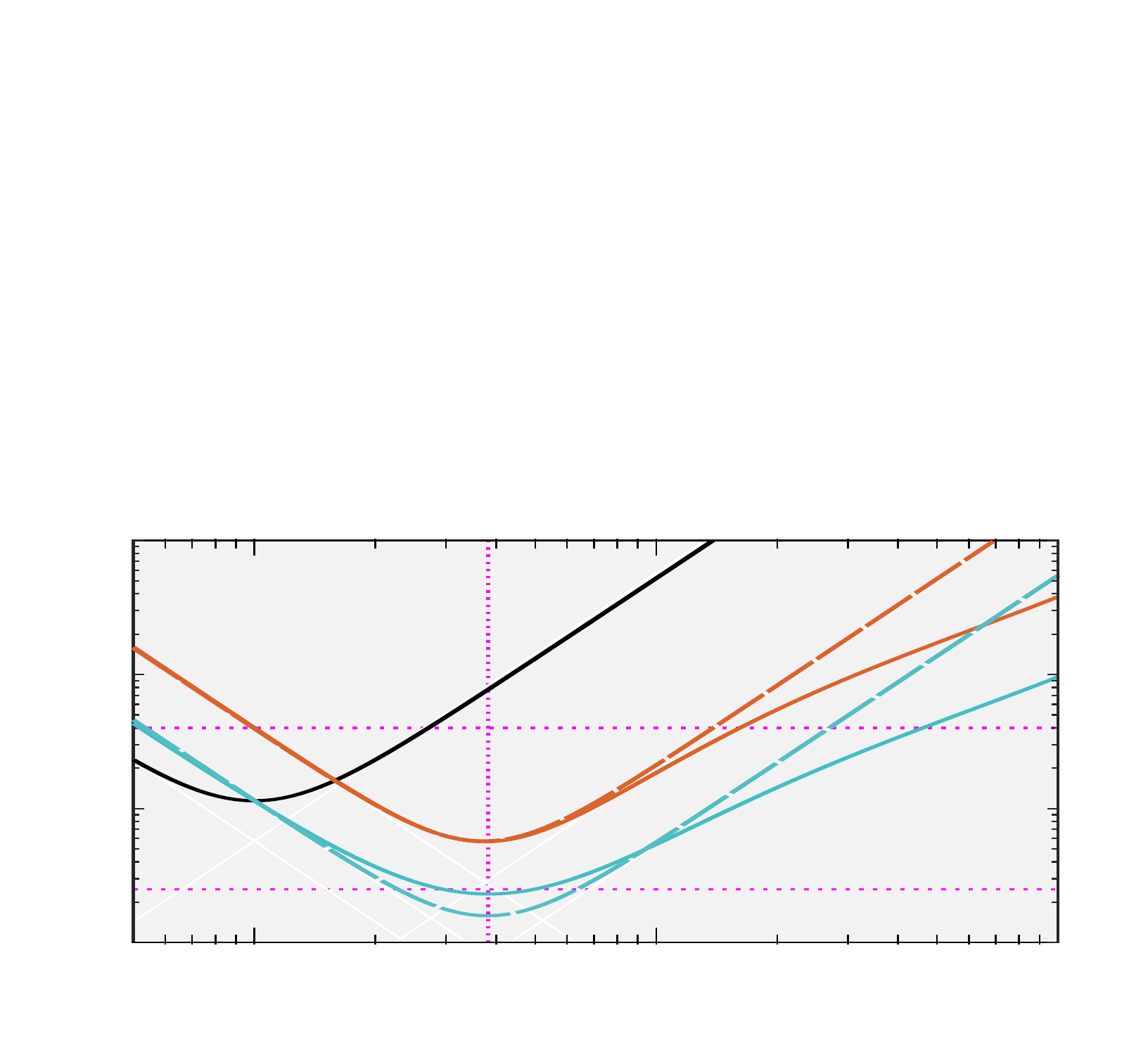
	\caption{\label{Fig:SurfaceOccupancy} Effective temperature of the oscillator at the endpoint of each cycle.\\
	\textit{a}) Effective cyclical steady-state occupancy $\n_{CSS}$ of the oscillator subject to perfect squeezing ($\epsilon = 0$; no cold bath) of magnitude $\mu$ at a rate of $\omega_{ap}$. White indicates $\n_{CSS} = \n_{H}$. Arrows indicate the directions of increasing $\n_{CSS}$. Other parameters are given in text.\\
	\textit{b}) $\n_{CSS}$ with imperfect squeezing. The cold bath occupancy is $\n_{C} = 2.5\times 10^{4}$, and $\epsilon$ has been adjusted at each $\omega_{ap}$ such that the effective $Q$ is held constant (\ie{} $\pi\omega_{M}/\epsilon\omega_{ap} = \omega_{M}/\gamma_{eff} = 10^{6}$).\\
		\textit{c}) Cross sections through parts \textit{a} (A, orange) and \textit{b} (B, blue) at $\omega_{ap}/\omega_{M} = 10^{3}$. The RWA result is given in black. In this parameter regime (with $\n_{C}$ less than an order of magnitude below $n_{H}$) there is a significant disagreement between the numerical results and the analytical approximation (dashed blue line). The analytical calculation of the optimum squeezing parameter, $\mu_{opt}$, is shown for case B (pink vertical line), whilst the pink horizontal lines indicate the hot and cold bath occupancies.}
\end{figure}

Representative calculations of $\n_{CSS}$ as a function of the squeezing strength $\mu$ and the squeezing application rate $\omega_{ap} = 2\pi/\tau$ are given in Fig.~\ref{Fig:SurfaceOccupancy} \textit{a}) \& \textit{b}). We use $\omega_{M} = 1$~MHz \& $Q = 10^{6}$---as might be expected for common micromechanical resonators such as SiN or SiC microstrings---and $\n_{H} = 4\times 10^{5}$ ($T_{H} \approx 300$~mK, achievable in a 3-He cryostat); we will continue to use these parameters unless otherwise stated. The remaining parameters are given in the figure caption.

Fig.~\ref{Fig:SurfaceOccupancy} shows that increasing $\mu$ (from $\mu = 1$) initially decreases $\n_{CSS}$, but at high $\mu$ the oscillator begins to heat once more. To understand this phenomenon we employed a combination of analytical and numerical techniques (see Appendix~\ref{App:SteadyState}), to find that
\begin{equation}
	\n_{CSS} \approx \frac{\Gamma \bar{n}_{H} + \gamma_{eff} \n_{C}}{\Gamma + \gamma_{eff}} \Br{\frac{1}{\mu^{2}} + \frac{\mu^{2}}{\mu_{opt}^{4}}},
	\label{Eqn:SteadyOccupancy}
\end{equation}
where $\gamma_{eff} \approx \epsilon\omega_{ap}/\pi$ is the effective decay rate to the cold bath, and $\mu_{opt}$ is the squeezing magnitude that optimizes (minimizes) the effective temperature. It can be shown that $\mu_{opt}$ is also equal to the amount of squeezing required to minimize the energy of the added noise (\ie{} minimize the trace of the added noise covariance $V_{add}$).

The functional form of \eqref{Eqn:SteadyOccupancy} can be traced to the competition between noises added to the position and momentum over the thermal cycle. As noted previously, the noise added from the hot bath predominantly affects the oscillator's momentum. Thus, in the absence of squeezing, $V_{add}$ is asymmetric. Let us imagine the system's behavior for increasing amounts of squeezing, beginning from $\mu =1$. For small increases in $\mu$, the squeezing reduces the asymmetry of $V_{add}$ and decreases the total noise energy ($\Tr{V_{add}}$) because $S_{2}$ attenuates the momentum noise. This behavior is linked to the first term ($\mu^{2}$) of \eqref{Eqn:SteadyOccupancy}. If $\mu$ is increased further the squeezing will reach an optimum value, $\mu_{opt}$, that minimizes $\n_{CSS}$; at this point the added noise is symmetric, minimising its energy. Beyond this point, any increase in $\mu$ begins to increase $\n_{CSS}$ again. This phenomenon, linked to the $\mu^{2}/\mu_{opt}^{4}$ term of \eqref{Eqn:SteadyState}, is due to amplification of the $X$ component of the added noise. Together, these effects yield a functional form that is reminiscent of the standard quantum limit for position measurement on a free mass, where an optimal interaction strength exists that balances measurement noise with quantum back-action noise \cite{BowenMilburn}.

\eqref{Eqn:SteadyOccupancy} is most accurate when $\Cu{Q,\n_{H}}\gg 1 \gg \Cu{\omega_{M}\tau,\epsilon}$. In this case we find
\begin{equation}
	\mu_{opt}^{4} \approx 3\Br{\frac{\omega_{ap}}{2\pi\omega_{M}}}^{2}\Sq{1+\frac{\gamma_{eff}\n_{C}}{2\Gamma\n_{H}}}.
	\label{Eqn:MuOpt}
\end{equation}
Together, \eqref{Eqn:SteadyOccupancy} and \eqref{Eqn:MuOpt} generally provide an excellent approximation to $\n_{CSS}$. As seen in Fig.~\ref{Fig:SurfaceOccupancy} \textit{c}), the approximation becomes less accurate as $\epsilon$ becomes larger.

Fig.~\ref{Fig:SurfaceOccupancy} also clearly shows that the squeeze--rotate--unsqueeze protocol can reduce the temperature of the oscillator to well below $\n_{H}$, opening up a new method of cooling mechanical oscillators to study their quantum behavior, and as a preparatory step for quantum sensing and information processing protocols \cite{BowenMilburn}. Furthermore, the oscillator may be made even colder than $\n_{C}$ during parts of the cycle. Although this is initially counter-intuitive, note that the working fluid of a conventional refrigerator is also reduced below the cold bath temperature for part of each cycle. We compare the cooling achieved using our scheme with that achieved by cold damping and sideband cooling---staple optomechanical cooling techniques---in Appendix~\ref{Sec:OtherCooling}.

\subsection*{Effective occupancy in the RWA} \label{Sec:EffectiveOccupancyRWA}

Repeating the same exercise in the RWA (for both hot and cold baths) yields
\begin{equation*}
	\upBr{\n_{CSS}}{\mathrm{RWA}} \approx \frac{\Gamma \bar{n}_{H} + \gamma_{eff} \n_{C}}{\Gamma + \gamma_{eff}} \cdot \half\Br{\frac{1}{\mu^{2}} + \mu^{2}},
\end{equation*}
which is smallest at $\mu = 1$ and never less than the simple detailed-balance equilibrium value of $\upBr{\n_{CSS}}{\mathrm{RWA}}|\sb{\mu=1} = \Br{\Gamma\n_{H}+\gamma_{eff}\n_{C}}\Br{\Gamma+\gamma_{eff}}^{-1}$. As shown in Fig.~\ref{Fig:SurfaceOccupancy}~panel~\textit{c}), this means that $\upBr{\n_{CSS}}{\mathrm{RWA}}$ is never less than the cold bath occupancy. This situation arises because the added noise is always symmetrical under the RWA; consequently, squeezing always increases the noise energy and the machine cannot work as a refrigerator.

\section*{Engine, pump, refrigerator} \label{Sec:Cycles}

As noted above, the squeeze--rotate--unsqueeze protocol is capable of reducing the oscillator's temperature to below that of both thermal baths. This makes it possible for the oscillator to act as the working fluid of a refrigerator. Similarly, the ability to reduce $\n_{CSS}$ to below $\n_{H}$---even when the cold bath is absent ($\epsilon = 0$, as in Fig.~\ref{Fig:SurfaceOccupancy}~panel~\textit{a}))---implies that the system can also behave as a heat pump.

In order to examine these behaviors quantitatively, we calculate the mean work ($W$), heat from the cold bath ($Q_{C}$), and heat from the hot bath ($Q_{H}$) during a cycle of evolution in the steady-state. A positive number indicates an influx of energy to the oscillator, and all results have been normalised to units of mechanical quanta.

It is clear that because the perfect squeezers $S\sb{j}$ are unitary (isentropic) they are associated with work ($W$), with the other operations corresponding to heat exchange. Put differently, the squeezer is driven by a classical field with a large amplitude, constituting a work source/sink, whilst all other interactions do not change the mean behavior of the baths (a characteristic of heat transfer). Thus
\begin{eqnarray}
		W & = & \frac{1}{4} \Tr{SV_{CSS}S^{\mathrm{T}} - V_{CSS}+S_{2}V^{\Br{3}}S_{2}^{\mathrm{T}}-V^{\Br{3}}}, \label{Eqn:W} \\
		Q_{H} & = & \frac{1}{4} \Tr{\upBr{V}{3}-\upBr{V}{2}}, \label{Eqn:QH} \\
		Q_{C} & = & -\Br{W+Q_{H}}, \label{Eqn:QC}
\end{eqnarray}
where $\upBr{V}{2}$ and $\upBr{V}{3}$ are the covariance matrices immediately after $S_{1}^{\prime}$ and immediately before $S_{2}^{\prime}$ respectively (\cf{} Fig.~\ref{Fig:SequenceSimplified} \textit{d})). \eqref{Eqn:QH} arises because of the first law of thermodynamics, \ie{} there is no net change in the system's energy over one full cycle.

\begin{figure}
	\centering
	\def\svgwidth{1\columnwidth}
	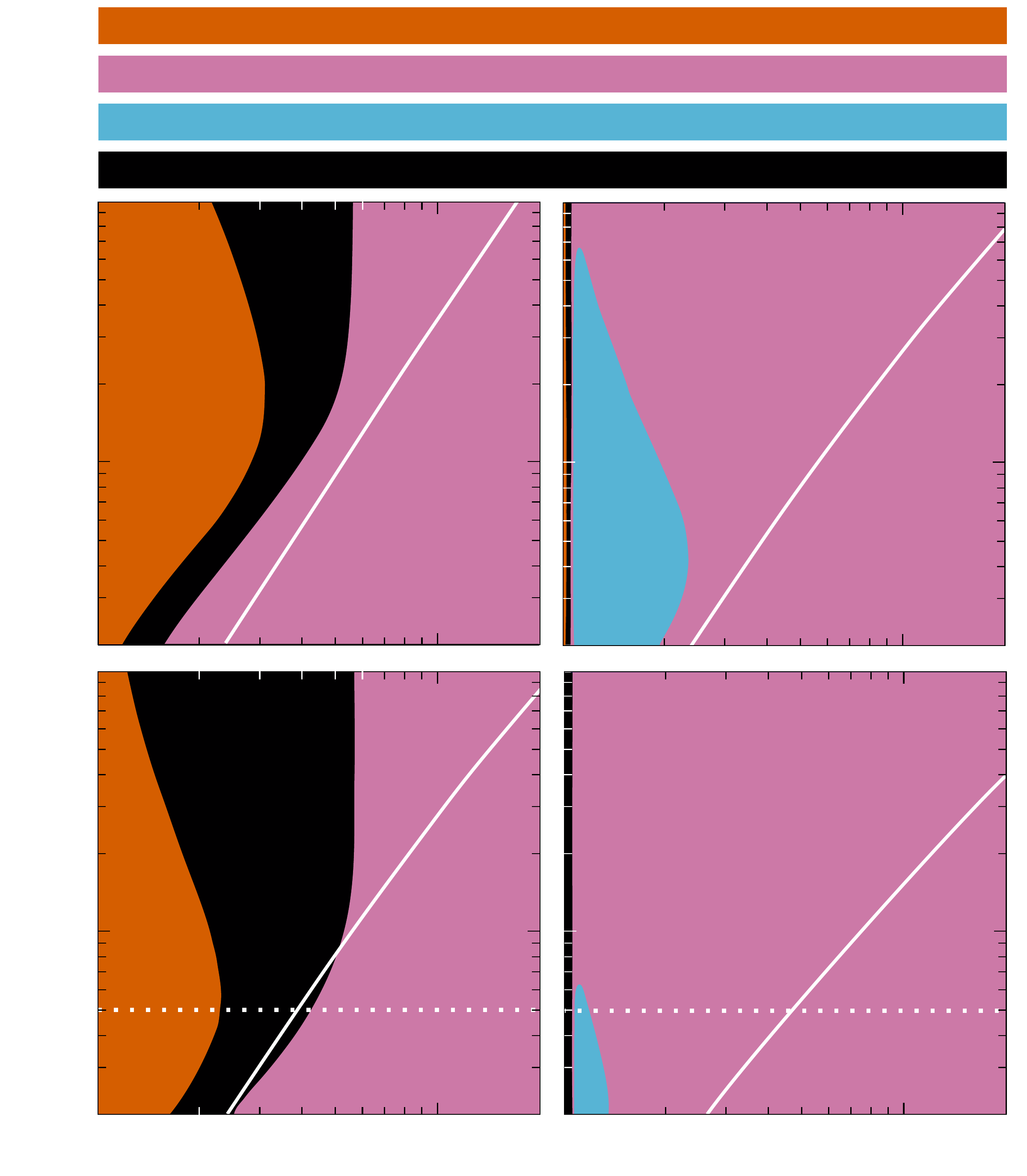
	\caption{\label{Fig:Phase} Operating regimes of the squeeze--rotate--unsqueeze protocol for different cold bath occupancies $\n_{C}$ and coupling constants $\epsilon$. In the black region energy flows from hot to cold even with the application of work. Other parameters given in text. The solid white lines show $\mu_{opt}$. Plots of thermodynamic efficiencies along the dashed white lines are given in Fig.~\ref{Fig:Efficiency}.
	}
\end{figure}

Numerical results show that the system can indeed act as a heat pump, fulfilling $Q_{H} < 0$ \& $W > 0$. It can also act as a refrigerator ($Q_{C} > 0$ \& $W > 0$), or a heat engine ($Q_{H} > 0$ \& $W < 0$). These regimes are shown in Fig.~\ref{Fig:Phase}. There is also a fourth `trivial' region in which the work performed on the system is positive but insufficient to reverse the flow of heat from the hot bath to the oscillator. Together, these form a parameter space of behaviors.

The heat pump behavior is straightforward to understand by considering the limit as the time between squeezers ($\tau$) tends to zero. The damping rate of the system into the hot bath then becomes $2 \Gamma \expect{P^{2}}$, where $\expect{P^{2}}$ is boosted above its CSS value by a factor of $\mu^{2}$ during the first squeezing interaction. Thus heat pumping occurs whenever this boosted loss rate overwhelms the noise coming in from the hot bath. This phenomenon is also behind the fact that the oscillator may be cooled to temperatures lower than that of the cold bath.

In the engine region the system is extracting work from the hot bath and dumping entropy into the cold bath. This occurs when the state $V^{\Br{3}}$ has a larger ratio of eigenvalues than $V^{\Br{2}}$. The momentum-damped Langevin equations permit this if
\begin{eqnarray}
	\n_{H} > \mu^{2}\n_{CSS} & \;\;\;(\mathrm{for}\;\; & \mu >1), \;\; \mathrm{or} \label{Eqn:WorkCriterion1} \\
	\n_{H} < \mu^{2}\n_{CSS} & \;\;\;(\mathrm{for}\;\; & \mu <1), \label{Eqn:WorkCriterion2}
\end{eqnarray}
assuming $\Cu{\epsilon,\epsilon\n_{C}} \ll 1 \ll \n_{H}$; see Appendix~\ref{App:EnginePhase}. For $\mu > 1$, this can be interpreted as requiring that noise enters from the hot bath faster than the system can damp into the hot bath (and \vv{} for $\mu < 1$).

Refrigeration---removing heat from the cold bath---only occurs when $\n_{C}$ is sufficiently large. The sign of $Q_{C}$ may be determined by considering the two loss steps involving the cold bath, as shown in Appendix~\ref{App:RefrigerationPhase}. For small $\epsilon$ we find the condition for refrigeration is
\begin{equation}
	\n_{C} > \frac{\n_{CSS}}{2}\Br{1 +\mu^{2}}.
	\label{Eqn:FridgeCondn}
\end{equation}

\subsection*{Operating regimes in the RWA} \label{Sec:PhasesRWA}

By explicitly calculating the steady-state covariance matrix in the RWA we were able to derive no-go theorems (Appendix~\ref{App:RefrigerationPhase}) that show that the heat engine and refrigerator regimes of operation are forbidden in the RWA. The former proof is valid in all parameter regimes satisfying basic requirements of physicality, whilst the latter is valid in the $\epsilon \ll 1$ and $\Gamma \ll \omega_{M} \ll \omega_{ap}$ regime considered throughout this paper. One therefore finds that there are only two operating regimes in the RWA: the `trivial' regime (where the input work and heat from the hot bath both flow into the cold bath), and a heat pump regime. Note that the heat pump simply dumps input work into the hot bath, because there can be no refrigeration.

\subsection*{Coefficients of performance} \label{Sec:COPs}

\begin{figure}
	\centering
	\def\svgwidth{1\columnwidth}
	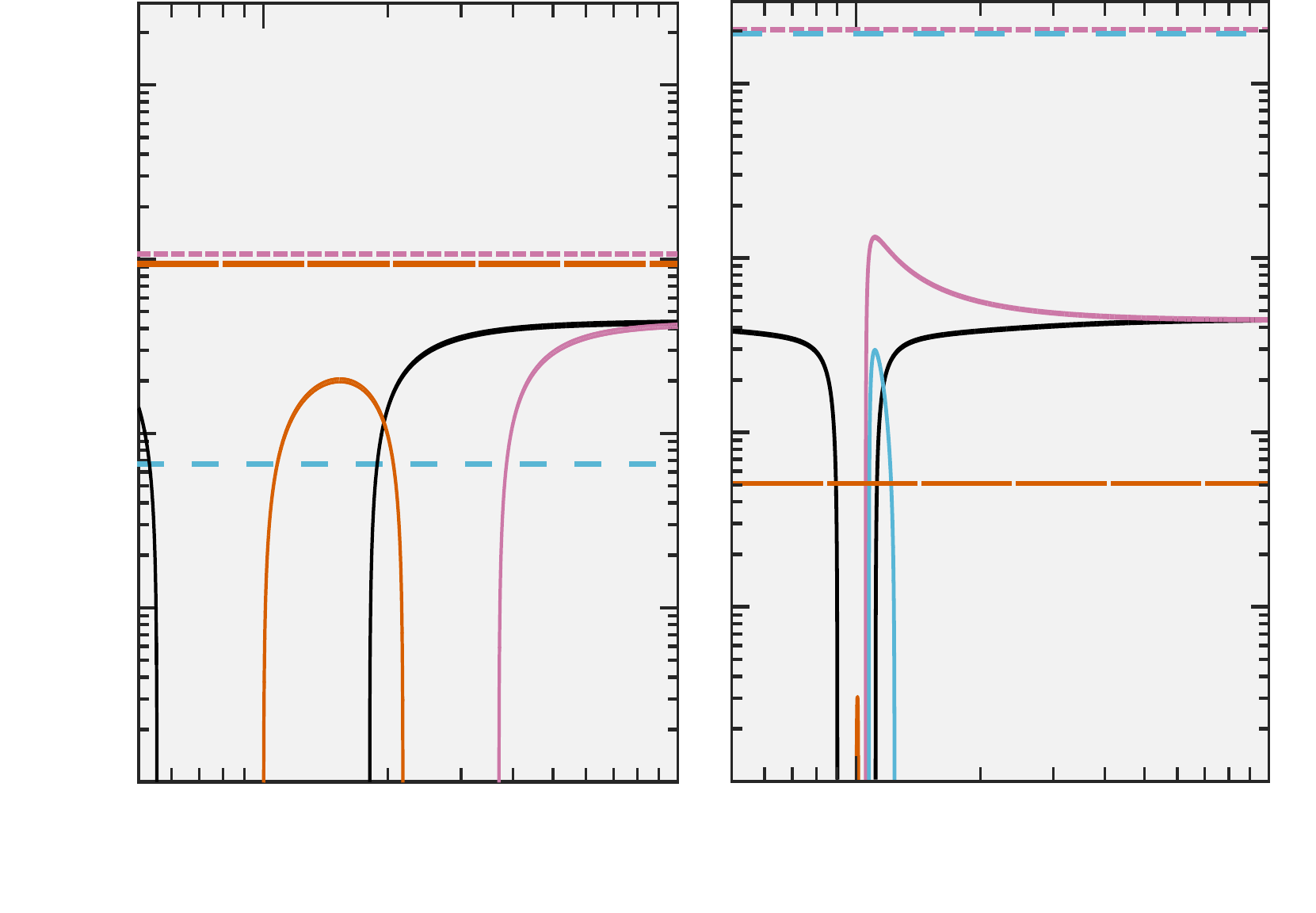
	\caption{\label{Fig:Efficiency} Coefficients of performance when operating as a heat pump (purple, P), heat engine (blue, E), or refrigerator (orange, R). The black line shows the RWA prediction for the heat pump regime. Thermodynamic bounds (upper limits) are shown as dashed lines (P, R, and E in order of increasing dash length). Panel \textit{a}) is a cross-section through Fig.~\ref{Fig:Phase}~\textit{a}) at $\omega_{ap}/\omega_{M} = 40$, and panel \textit{b}) shows the corresponding section of Fig.~\ref{Fig:Phase}~\textit{b}). 
	$Q = 20$, $n_{H} = 4\times 10^{5}$, $\epsilon = 10^{-2}$: other system parameters are given in the main text.}
\end{figure}

Finally, we can consider the performance of the available thermodynamic cycles. The relevant coefficients of performance are defined by
\begin{eqnarray}
	\COP_{pump}   & = & \mod{Q_{H}/W} \leq \eta^{-1}, \label{Eqn:COPPump}\\
	\COP_{engine} & = & \mod{W/Q_{H}} \leq \eta, \label{Eqn:COPEngine}\\
	\COP_{fridge} & = & \mod{Q_{C}/W} \leq \Br{1-\eta}/\eta, \label{Eqn:COPFridge}
\end{eqnarray}
where the thermodynamic limits are expressed in terms of the Carnot efficiency, $\eta = 1-T_{C}/T_{H}$.

Representative calculations are provided for cross-sections through Fig.~\ref{Fig:Phase}~\textit{b}) and~\textit{d}), as shown in Fig.~\ref{Fig:Efficiency}. They demonstrate that (for these parameters) the peak engine efficiency is approximately $18\%$ of the Carnot limit. They also confirm that the maximum heat pumping efficiency ($\sim 1.13$) is achieved when the squeezing strength $\mu$ is below the value $\mu_{opt}$ that minimizes the steady-state occupancy. Conversely, in the case that both baths are treated using the RWA, the heat pump efficiency improves monotonically with $\mu$ (for $\mu > 1$), and does not exceed unity. This difference arises because the RWA does not have a refrigerating regime (nor an engine regime) as a result of the lack of asymmetry in the added noise. These conclusions are essentially unchanged in the case that the RWA is only made on the hot bath, so long as $\n_{H} \gg \n_{C}$.

\section*{Experimental feasibility} \label{Sec:Feasibility}

This manuscript is primarily concerned with presenting the concept of the squeeze--rotate--unsqueeze protocol and examining the underlying physical mechanisms. However, we ought to give a basic overview of the experimental requirements for implementing such a scheme. We will begin by explicitly linking the model of squeezing used here to the pulsed optomechanical squeezer of \cite{Bennett2018}; then consider the limitations of the model used in this work; examine its performance with specific examples \cite{Peterson2019,Wilson2015}; and summarise future work arising from this research.

\subsection*{Connection with pulsed optomechanics} \label{Sec:RealSqueezer}

The simple imperfect squeezer presented above can be derived as a limiting case of the rapid pulsed optomechanical squeezing scheme proposed in \cite{Bennett2018}. In this scheme, a sub-mechanical-period optical pulse interacts with the mechanical oscillator four times, with variable time delays between each interaction. Careful choice of the input pulse's squeezing, time delays, and optomechanical interaction strengths makes it possible to generate high-purity squeezing of the mechanical state.

The amount of mechanical squeezing is described by $\mu = \Br{1-\chi_{1}^{2}\tan\varphi}^{-1}$, where the first pulsed interaction strength is $\chi_{1}$ and $\varphi$ is a mechanical rotation angle that must be inserted between pulses two and three, corresponding to a finite delay time. The optomechanical interaction strength depends on the single-photon optomechanical coupling rate $g_{0}$, mean photon number in the pulse $N$, and the cavity linewidth $\kappa$, \viz{} $\chi = -8g_{0}\sqrt{N}/\kappa$. The other interaction strengths are fixed by the choice of $\chi_{1}$ and $\varphi$.

In the absence of decoherence, the transformation generated by the total sequence of four pulses is
\begin{eqnarray}
		X & \rightarrow & X/\mu + \Br{1-\mu}\tan\varphi P \nonumber \\
		 & & {}+ \sqrt{2\mod{1-\mu^{-1}}\tan\varphi} X_{L}^{\Br{\phi_{opt}}}, \label{Eqn:XOpt} \\
	P & \rightarrow & \mu P, \label{Eqn:POpt}
\end{eqnarray}
where $X_{L}^{\Br{\phi_{opt}}}$ is optical noise coupled in from the pulses. Thus, we see that the ideal squeezer can be achieved in the limit that the pulse is highly squeezed, \viz{}
\[
	\expect{ \Br{X_{L}^{\Br{\phi_{opt}}}}^{2}} \rightarrow 0,
\]
and $\varphi\rightarrow 0$ (with the caveat that $\chi_{1}$ is simultaneously boosted, in order to keep $\mu$ fixed). In the practical case of finite $\varphi$ (and context of this thermal machine) we require that $\varphi \ll 2\pi\omega_{M}/\omega_{ap}$, so that $S_{1}^{\prime}$ is fast compared to the hot bath interaction time $\tau$. Further details on these requirements are given in below.

It is also important to consider the effects of noise added during the pulsed squeezer. There are two relevant loss mechanisms; damping into the mechanical bath (during the delay $\varphi$), and loss on the optical ancilla (\cf{} \cite{Bennett2018}). The former can be treated using the viscous damping model already discussed. The optical loss can be modeled by a beamsplitter interaction between the optical ancilla and a thermal mode of occupancy $\n_{C}$, with a reflectivity of $\epsilon_{BS}$. It is appropriate to use a beamsplitter interaction---\ie{} make the RWA---in this situation because of the high optical (or microwave) frequencies involved. This optical noise does not immediately affect the mechanical oscillator, but is instead coupled into it during the last two pulsed interactions with the oscillator. By the conclusion of this sequence the optical noise
\[
	V_{add} = \diag{0, \; 4\epsilon_{BS}\Br{\mu-1}\mu\csc 2\varphi\cdot \Br{2\n_{C}+1}}
\]
has been added onto the mechanical oscillator. Note that no noise couples into position, exactly as would be expected from an interaction with a bath through viscous damping. In the parameter regime of interest---small optical loss ($\epsilon_{BS}$), small to moderate $\mu$, and short rotation times---this can be interpreted as a realisation of the noise given by \eqref{Eqn:VC}. We then simply need to ensure that the mechanical bath adds much less noise than the optical bath, \ie{}
\[
		\varphi \ll \frac{\epsilon\Br{2\n_{C}+1}}{Q^{-1}\Br{2\n_{H}+1}}.
\]
In this limit the aggregate effect of decoherence during the squeezer can be modeled by viscous damping into a bath with occupancy $\n_{C}$ and loss $\epsilon = 1-\sqrt{1-4\epsilon_{BS}\Br{\mu-1}\mu\csc 2\varphi}$.

We can conclude that, within the limits outlined above, our imperfect squeezer model is both simple to use and a reasonable representation of the full protocol presented in \cite{Bennett2018}.

Note that in this section we have only considered the first squeezer, $S_{1}^{\prime}$. Realising the second squeezer using pulsed optomechanics requires an extended sequence of \textit{six} pulses because of the quadrature angle at which it must operate ($2\pi\omega_{M}/\omega_{ap} \neq 0$). Numerical calculations indicate that this can be achieved with a fidelity exceeding $95\%$ for moderate squeezing parameters; further details may be found in Appendix~\ref{App:SixPulse}.

\subsection*{Requirements for validity of model} \label{Sec:ModelValidity}

The simplified theory employed in this work is valid in the regime where
\begin{equation}
	\kappa \gg \omega_{BW} \gg \omega_{sq} \gg \omega_{ap} \gg \omega_{M} \gg \Cu{g_{0},\Gamma},
	\label{Eqn:OperatingRegime}
\end{equation}
with $\omega_{BW}$ being the optical pulse bandwidth and $\omega_{sq} = 2\pi\omega_{M}/\varphi$ being the rate associated with the short delay between pulses two and three during the squeezer. This hierarchy of frequencies arises as follows---from the squeezer \cite{Bennett2018} itself we have that the single-photon optomechanical coupling must be weak ($g_{0} \ll \Cu{\omega_{M},\kappa}$); the mechanical quality high ($\Gamma \ll \omega_{M}$); the pulse much shorter than the mechanical period ($\omega_{M} \ll \omega_{BW}$); and the cavity must respond adiabatically to the input field ($ \omega_{BW} \ll \kappa$). In addition, our treatment of the heat machine requires that the pulses used to construct the squeezers be much faster than the squeezer delay time ($\omega_{sq}^{-1}$) and the time spent in contact with the hot bath ($\omega_{ap}^{-1}$), which translates to $\omega_{BW} \gg \omega_{sq} \gg \omega_{ap}$.

It is possible to use exponentially-rising pulses to efficiently and rapidly couple light into the optical cavity \cite{Bader2013}, relaxing the $\kappa \gg \omega_{BW}$ requirement to $\kappa \sim \omega_{BW}$. This reduces the optical power required by approximately one hundred (as seen from \eqref{Eqn:PhotonNumber}) at the expense of adding experimental complexity in the form of requiring pulse reshaping between each QND interaction. Reshaping could be accomplished using methods such as those presented in \cite{Andrews2015}. Taking advantage of these techniques relaxes \eqref{Eqn:OperatingRegime} to
\begin{equation}
	\kappa \sim \omega_{BW} \gg \omega_{sq} \gg \omega_{ap} \gg \omega_{M} \gg \Cu{g_{0},\Gamma}.
	\label{Eqn:OperatingRegimeRelaxed}
\end{equation}

\subsubsection*{Photon number} \label{Sec:PhotonNumber}

One direct consequence of \eqref{Eqn:OperatingRegime} is that the number of photons required for the protocol must always be fairly large (for $\mu \neq 1$). To see this, we first inspect Eq.~(21) of \cite{Bennett2018}, which gives the total photon number required to perform $S_{1}$ (valid for small $\varphi$), \viz{}
\begin{equation}
	N_{tot,1} = \frac{1}{\tan\varphi}\Br{\frac{\kappa}{8 g_{0}}}^{2}\mod{1-\frac{1}{\mu}}\Br{3+\frac{1+\Br{1-\mu^{-1}}^{2}}{\mu^{2}}}.
	\label{Eqn:PhotonNumber}
\end{equation}
Note that the photon number grows as $\Br{\kappa/g_{0}}^{2}$. From \eqref{Eqn:OperatingRegime}, we see that $\kappa/g_{0}$ must necessarily be a large number, not less than approximately $10^{5}$ even for quite considerable single-photon coupling rates. This can be relaxed to $\sim 10^{4}$ by using pulse shaping techniques to reduce the required pulse bandwidth and cavity linewidth, as in \eqref{Eqn:OperatingRegimeRelaxed}. The photon number is also proportional to $\cot\varphi \approx \varphi^{-1} \gg 1$, boosting this prefactor even higher. We can therefore see that an experimental realisation of our thermal machine using the approach identified here would need to be \textit{a}) close to the $g_{0}\sim\omega_{M}/10$ regime, so as to obtain the largest values of $g_{0}$ compatible with \eqref{Eqn:OperatingRegimeRelaxed}; and \textit{b}) capable of handling high optical powers. Low $\omega_{M}$ is also desirable because of the practical demands of manipulating the optical pulses (rotating and potentially reshaping them) during each squeezer \cite{Bennett2018}.

Let us consider two exemplary systems; the superconducting three-dimensional cavity electromechanical system detailed in \cite{Peterson2019}, and the evanescently-coupled optomechanical resonator described in \cite{Wilson2015}. Relevant parameters for these are given in Table~\ref{Tab:Params}. For a modest $3$~dB of squeezing ($\mu = \sqrt{2}$) and $\omega_{ap}/\omega_{M} = 10$ these require a number of photons of order $\sim 10^{13}$ (electromechanical) and $\sim 10^{10}$ (optomechanical) to complete $S_{1}$. These correspond to time-averaged input powers of $1.5$~mW and $3.3$~W, respectively (bearing in mind that $S_{2}$ will require a similar number of photons, \cf{} Appendix~\ref{App:SixPulse}). These are large but not entirely unreasonable average input powers. For example, the electromechanical system has been demonstrated to handle a steady-state input power of up to $13~\upmu$W ($10^{9}$ intracavity photons), with this upper limit enforced by optomechanical instabilities rather than damage \cite{Peterson2019}.

We can also estimate the peak input powers, based on the pulse bandwidths, yielding $450$~mW and $1$~kW inputs respectively. These indicate that the optomechanical cavity may not be capable of withstanding the peak pulse irradiance without damage at the glass--air interface; the estimated peak irradiance is on the order of $1$~TW/cm\textsuperscript{2}, which exceeds the damage threshold by a factor of $10^{3}$ (\eg{} \cite{Koechner1988}, chapter 11 thereof).

Other potential platforms have desirable $g_{0}/\omega_{M}$ ratios\footnote{\cite{Baker2016} and \cite{Murch2008} actually exceed a ratio of unity, but this could easily be reduced.}, but are also likely to be incapable of handling the required input power, \eg{} superfluid helium thin-film optomechanics ($g_{0}\sim\omega_{M}$) \cite{Baker2016}, ultracold atom optomechanics ($g_{0}\sim\omega_{M}$) \cite{Murch2008}, and optomechanical trampoline resonators ($g_{0}\sim10^{-2}\omega_{M}$) \cite{Kleckner2011}. Consequently, an electromechanical implementation \ala{} \cite{Peterson2019} appears most practical at this time. As shown in Table~\ref{Tab:Params}, it is likely that this electromechanical system \cite{Peterson2019} could be shifted into a practical power-handling regime by at most a ten-fold increase in $g_{0}$. For comparison, at least a thirty-fold increase in $g_{0}$ is required to bring the optical system \cite{Wilson2015} into the $\lesssim 1$~GW/cm\textsuperscript{2} regime in which optical damage is expected to be manageable.

Finally, it is worth noting that achieving a level of squeezing sufficient to quantum squeeze the oscillator (\cf{} Fig.~\ref{Fig:TimeDependence}) would only require a small increase in the photon number above the levels described here; approximately a factor of three (at lower temperatures, \cf{} Fig.~\ref{Fig:TimeDependence}).

\begin{table}
\begin{center}
\begin{tabular}{l | ll | ll}
 & \cite{Peterson2019} & {} & \cite{Wilson2015} \\
 & Microwave & (Upgraded) & Optical & (Upgraded) \\
\hline
$\omega_{M}/2\pi$ & 970~kHz\textsuperscript{*} & {} & 4.3~MHz & {} \\
$g_{0}/2\pi$ & 167~Hz & 1.67~kHz & 20~kHz & $600$~kHz \\
$\omega_{cav}/2\pi$ & 6.5~GHz & {} & 2.4~PHz & {} \\
$Q_{M}$ & $3\times 10^{5}$ & {} & $7.6\times 10^{5}$ & {} \\
\hline
$\omega_{ap}/2\pi$ & 9.7~MHz & {} & 43~MHz & {} \\
$\omega_{BW}/2\pi$ & 970~MHz & {} & 4.3~GHz & {} \\
$\kappa/2\pi\textsuperscript{**}$ & 1.9~GHz & {} & 8.6~GHz & {} \\
\hline
$N_{tot,1}$ & $3.5\times 10^{13}$ & $3.5\times 10^{11}$ & $4.8\times 10^{10}$ & $5.3\times 10^{7}$
\end{tabular}
\end{center}
\caption{\label{Tab:Params} Exemplary parameters for state-of-the-art electro- and opto-mechanical systems. The target amount of squeezing is 3~dB ($\mu = \sqrt{2}$), with a modest delay of $\varphi/2\pi = 0.01$. System parameters are as given in the first block of the table (except as noted below). Sample application rates, pulse bandwidths, and optical linewidths are in the second block; they obey \eqref{Eqn:OperatingRegimeRelaxed}. The final line shows the total photon number required for the first squeezer in each thermal cycle, $N_{tot,1}$. We have also included `upgraded' parameter sets showing suggested improvements required for the scheme to be experimentally feasible.\\
\textsuperscript{*} this has been reduced by $10$ so that $\kappa < \omega_{cav}$ is satisfied.\\
\textsuperscript{**} strongly overcoupled to satisfy \eqref{Eqn:OperatingRegimeRelaxed}
}
\end{table}

\subsubsection*{Intrinsic imperfections}

There are a two undesirable non-idealities---appearing in \eqref{Eqn:XOpt} even in the absence of loss---that can also ruin the squeezer's performance; one adds a component of $P$ into the transformation (that should ideally be diagonal), and the other adds noise from the (squeezed) optical ancilla. As already mentioned, both terms may be suppressed by operating with extremely small $\varphi$. As seen above, doing so requires that the optical pulse strengths be increased to keep $\mu$ fixed, according to \eqref{Eqn:PhotonNumber}.

To show that these terms are negligible in most regimes, let us consider their effect on the fidelity of preparation of a squeezed thermal state using the parameters described above. Suppose that the initial occupancy of the oscillator is $\n_{eff} = 10^{3}$; it then undergoes $S_{1}^{\prime}$ with $\omega_{M}/2\pi = 970$~kHz and $g_{0}/2\pi = 167$~Hz. In the absence of damping, the fidelity of the actual output with the desired output is $0.999661$, even with an unsqueezed (vacuum) ancilla state. An ancilla with $10$~dB of squeezing yields a marginal improvement to a fidelity of $0.999662$. We can conclude that these unwanted terms are therefore not important when modelling our protocol (in the limit described); they are sufficiently small compared to the desired terms.

\subsubsection*{Extrinsic imperfections---displacement operations}

The optomechanical squeezer \cite{Bennett2018} requires that the optical field (quadratures) undergo set displacements (rotations) between each interaction with the mechanical oscillator. This could be achieved by interfering the pulse with a very bright control pulse on a highly asymmetric beamsplitter; the noise on the ancilla remains essentially unchanged, but the coherent amplitude is altered. Though simple to explain and implement, this approach is infeasible for our squeeze--rotate--unsqueeze machine at low $\epsilon$ (\eg{} $\epsilon < 10^{-1}$) because it requires impractically powerful control pulses, well above material damage thresholds.

\begin{figure}
	\centering
	\def\svgwidth{1\columnwidth}
	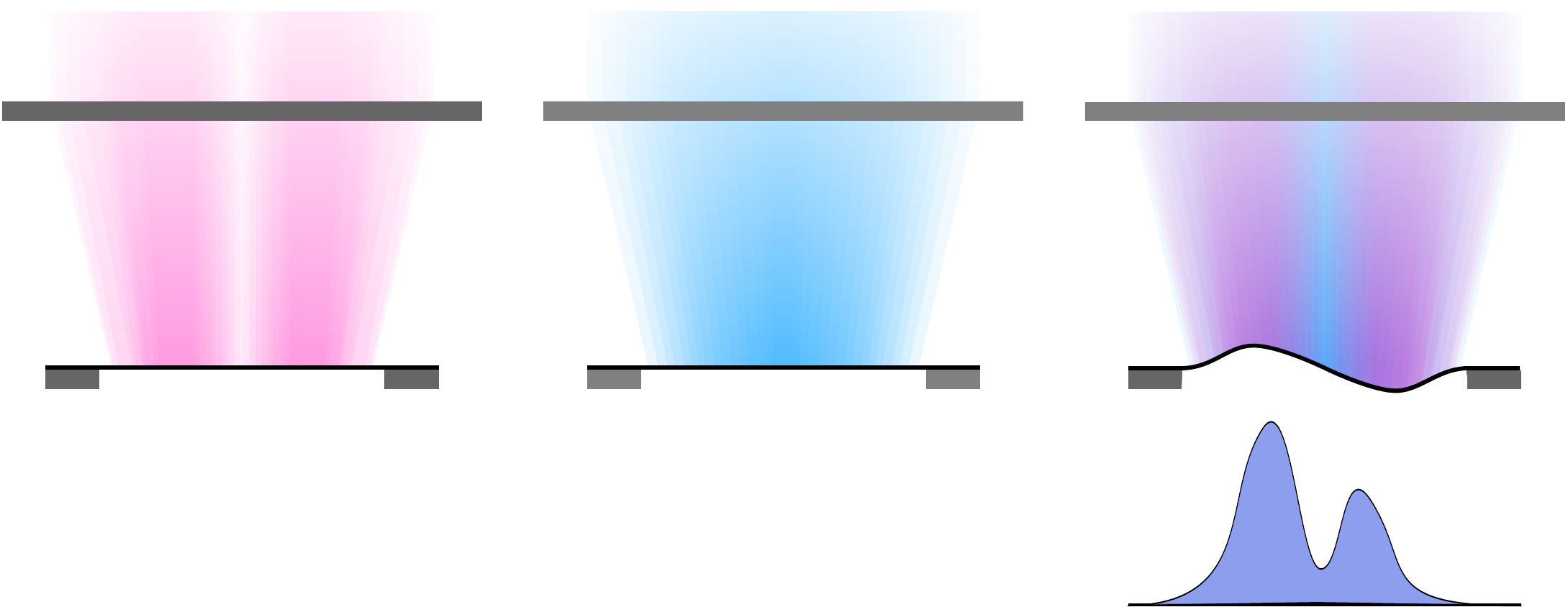
	\caption{\label{Fig:Interfere} Illustration of two-mode pulsed optomechanical scheme. The optomechanical cavity is composed of a rigid mirror (top) and flexible membrane (bottom).\\
	\textit{a}) Odd `science' mode, $a$. The intensity distribution (below) is an even function of the transverse coordinate ($x$), and will not excite odd mechanical modes of the membrane.\\
	\textit{b}) Even `control' mode, $b$. Again, the intensity is even and will not excite odd mechanical modes.\\
	\textit{c}) The beating of $a$ and $b$ causes a time-dependent intensity distribution that can drive antisymmetric mechanical modes.}
\end{figure}

This problem could be circumvented by moving to an interferometric scheme in which two optical modes are used to drive the mechanical oscillator. Consider the simple case of a \FP{} cavity where one mirror is composed of a flexible membrane, as in Fig.~\ref{Fig:Interfere}. The excitation of any individual transverse optical mode will not excite an odd mechanical mode, because their intensity distributions are all even. However, when two optical modes of opposite parity are excited simultaneously, they interfere and produce an odd beat-note that can drive an odd mechanical mode. The Hamiltonian describing this interaction is
\begin{equation}
	H = \hbar \upBr{g_{0}}{2} \Br{ a^{\dagger}b + b^{\dagger}a } X,
\end{equation}
where $a$ ($b$) is the odd (even) optical mode annihilation operator, $X$ is the position of the odd mechanical mode, and the coupling rate $\upBr{g_{0}}{2}$ depends on the exact geometry of the cavity. Let us designate $a$ as the `science' mode, and $b$ as the `control' mode. If $b$ is driven into a bright coherent state $\beta$ with phase $\phi$ (such that $\mod{\beta} = \mod{\langle b\rangle} \gg \mod{\langle a \rangle}$), then the Hamiltonian becomes
\begin{eqnarray}
	H & \approx & \hbar \upBr{g_{0}}{2} \mod{\beta\td} \Br{ a^{\dagger}\e^{\I\phi} + a \e^{-\I\phi}} X \nonumber \\
	& = & \hbar \upBr{g_{0}}{2} \mod{\beta\td} X_{a}\Br{\phi} X,
\end{eqnarray}
where $X_{a}\Br{\phi}$ is the $\phi$ quadrature of the science mode. Thus, we can control the quadrature that the oscillator interacts with simply by changing the phase of the control field.

One can imagine implementing this in a variety of ways. For example, the control mode and science mode could both be driven by pulsed inputs, with only the science mode being recirculated between each step of the squeezer. It may also be possible to make the science mode significantly higher-$Q$ than the control mode, so that the squeezed optical ancilla can be retained in the cavity whilst the optomechanical interaction is switched on and off using control pulses. Alternatively, this scheme could be generalized to make use of multiple \textit{longitudinal} modes in a long optical cavity, such that the field inside the cavity constitutes a pulse that bounces from one mirror to the other; all time delays in the system could then be locked to multiples of the cavity round-trip time. Each of these approaches has potential pitfalls and complications, full analyses of which lay outside the scope of this paper.

\subsection*{Feasibility---closing remarks} \label{Sec:ClosingRemarks}

We have seen that the protocol we have introduced in this paper is potentially realizable using modern electromechanical platforms, with high demands on peak circulating power and optical quality factor.

There are a substantial number of questions remaining to investigate, but they are outside the scope of this treatment. For instance, future work could examining the effects of finite-time squeezing operations, optimize the arbitrary-angle squeezer (Appendix~\ref{App:SixPulse}), investigate the effects of classical noise (such as pulse timing and amplitude variations), or construct a full numerical model including all decoherence and noise channels, \etc{} It is likely that pursuing these goals will reveal methods by which the experimental requirements can be relaxed further; our calculations throughout this paper have taken many limiting cases in order to keep the theory analytically tractable and the physical interpretation of our findings transparent, so are likely to be sub-optimal for a direct implementation.

It may also be possible to explore non-optomechanical implementations of our protocol in other systems that permit pulsed quantum non-demolition interactions, such as atomic spin ensembles \cite{Kohler2017}, \eg{} recent work \cite{Kohler2017} with effective coupling rates satisfying $g_{0}/\omega_{M}$ of $\sim 2\times 10^{-3}$.

\section*{Conclusion} \label{Sec:Conclusion}

We have proposed and modeled a thermodynamic system based on a momentum-damped mechanical oscillator subjected to rapid squeezing operations. Our calculations (code available on request) indicate that such a system can operate as a heat pump, a refrigerator, or a heat engine. Importantly, if lossy evolution is modeled by the BMME the latter two effects vanish. This indicates the emergence of rich---and potentially useful---quantum thermodynamical phenomena beyond the RWA. As it stands, our proposal is likely to be realizable with state-of-the-art electromechanical systems, but high power-handling capacity is necessary. Further optimization of the protocol may relax experimental requirements.

\section*{Acknowledgments} \label{Sec:Ack}

The authors thank Mr Stefan Forstner and Dr Kiran Khosla for discussions concerning pulsed optomechanics, and Prof. Ronnie Kosloff for discussion of Markovian Langevin equations. This work was funded by the Australian Research Council (ARC), CE110001013. JSB was supported by an Australian Government Research Training Program Scholarship. WPB acknowledges an ARC Future Fellow (FT140100650).

\section*{References} \label{Sec:References}

\appendix
\onecolumn
\pagebreak

\section*{Supplemental Material}

This Supplemental Material discusses the;
\begin{itemize}
	\item outline of the relation between the independent oscillator model and BMME (Appendix~\ref{App:BMMEandLOM});
	\item detailed dynamics of the oscillator under both approximations (Appendix~\ref{App:HarmonicOscillator});
	\item derivation of the steady-state occupancy (Appendix~\ref{App:SteadyState});
	\item comparison to dedicated optomechanical cooling procedures (Appendix~\ref{Sec:OtherCooling});
	\item derivation of the bounds on thermodynamic cycles in the independent oscillator model (Appendix~\ref{App:ThermoCycles});
	\item no-go theorems for heat engines and refrigerators in the RWA (Appendix~\ref{App:NoGo}); and
	\item six-pulse optomechanical sequence for an optomechanical squeezer at an arbitrary angle (Appendix~\ref{App:SixPulse}).
\end{itemize}

\section{Limitations of Markovian Langevin equation} \label{App:BMMEandLOM}

The Langevin equations, \eqref{Eqn:Xdot} and \eqref{Eqn:Pdot}, can be derived from the independent oscillator model---the most general microscopic model of a linear, passive heat bath \cite{SuppFord1997,SuppFord1988}. One must formally solve the equations of motion for the bath and back-substitute them into the system's. Then, in order to obtain Markovian equations, one must assume a `high' bath temperature and continuous bath spectrum, then neglect `initial slip' terms (which contribute only at time nought and order $Q^{-1}$). A full derivation is given by Giovannetti and Vitali \cite{SuppGiovannetti2001}. The results are not of Lindblad form \cite{SuppLindblad1976, SuppKohen1997} and are only valid when $\n_{H}$ and $\n_{C}$ are sufficiently large.

\section{Damped harmonic motion} \label{App:HarmonicOscillator}

\subsection*{Independent oscillator model} \label{Sec:IndependentOscillator}

Consider the equations of motion of a harmonic oscillator of (angular) frequency $\omega_{M}$ subject to momentum-dependent damping at rate $\Gamma$, \eqref{Eqn:Xdot} and \eqref{Eqn:Pdot}. Note that the loss appears only in \eqref{Eqn:Pdot}, as does the noise operator $\hat{\xi}\td$. These equations are typically employed to describe classical systems, but may be readily---if not straightforwardly---quantized (\eg{} \cite{SuppGardinerZoller}).

Integrating the equations of motion is straightforward because of their linearity. One obtains
\begin{eqnarray}
		X\td & = & \e^{-\Gamma t/2} \Sq{ \cos\Br{\sigma\omega_{M}t} + \frac{\Gamma}{2 \sigma \omega_{M}}\sin\Br{\sigma\omega_{M}t} }X\Br{0} +\e^{-\Gamma t/2} \frac{1}{\sigma}\sin\Br{\sigma\omega_{M}t}P\Br{0} +\delta X\td, \label{Eqn:XSoln} \\
		P\td & = & \e^{-\Gamma t/2} \Sq{ \cos\Br{\sigma\omega_{M}t} - \frac{\Gamma}{2 \sigma \omega_{M}}\sin\Br{\sigma\omega_{M}t} }P\Br{0}-\e^{-\Gamma t/2} \frac{1}{\sigma}\sin\Br{\sigma\omega_{M}t}X\Br{0} +\delta P\td. \label{Eqn:PSoln}
\end{eqnarray}
Here the modified resonance frequency is $\sigma \omega_{M}$ with $\sigma = \sqrt{1-\Gamma^{2}/4\omega_{M}^{2}}$ (we have assumed an underdamped oscillator, $\Gamma < 2 \omega_{M}$), and the effect of the thermal noise is captured by the increments $\delta X$ and $\delta P$. The solutions \eqref{Eqn:XSoln} and \eqref{Eqn:PSoln} may be recast as a matrix equation of the form
\begin{equation}
	\bm{X}\td = M_{H} \td \bm{X}\Br{0} + \bm{F}\Br{t},
	\label{Eqn:MatrixEOMs}
\end{equation}
where
\[
	M_{H} = \e^{-\frac{\Gamma t}{2}}
			\Br{\begin{array}{c c}
			\cos\Br{\sigma\omega_{M}t} + \frac{\Gamma}{2 \sigma \omega_{M}}\sin\Br{\sigma\omega_{M}t} & \frac{1}{\sigma}\sin\Br{\sigma\omega_{M}t} \\
			-\frac{1}{\sigma}\sin\Br{\sigma\omega_{M}t} & \cos\Br{\sigma\omega_{M}t} - \frac{\Gamma}{2 \sigma \omega_{M}}\sin\Br{\sigma\omega_{M}t}
			\end{array}}
\]
and
\begin{equation}
	\bm{F}\Br{t} = \Br{\begin{array}{c} \delta X\td \\ \delta P\td \end{array}} = \sqrt{2\Gamma} \intfin{t^{\prime}}{0}{t} M\Br{t-t^{\prime}} \Br{\begin{array}{c} 0 \\ \hat{\xi}\Br{t^{\prime}}\end{array}},
	\label{Eqn:Noise}
\end{equation}
with $\bm{X} = \trans{\Br{X \; P}}$.

In the high-temperature, high-$Q$ limit the noise operator $\hat{\xi}\td$ becomes Markovian and Gaussian. We shall henceforth restrict our attention to this case. The noise correlation function becomes
\[
	\mathcal{N}\Br{t,t^{\prime}} = \frac{\expect{\xi\td\xi\Br{t^{\prime}} + \xi\Br{t^{\prime}}\xi\td}}{2} = \Br{2 \n_{H}+1}\delta\Br{t-t^{\prime}}.
\]

Gaussian noise and quadratic operations (\ie{} squeezing, rotations, linear loss, and beamsplitters) imply that the steady-state will be characterized by its first and second moments. It is thus appropriate to consider the covariance matrix, given by
\begin{equation}
	V = \frac{1}{2}\Cu{ \expect{\bm{X}\trans{\bm{X}}} + \trans{\expect{\bm{X}\trans{\bm{X}}}} }.
\end{equation}
From Supp.~\eqref{Eqn:MatrixEOMs} we obtain the solution 
\begin{equation}
	V\td = M\td V\Br{0} \trans{M}\td + V_{H}\Br{t},
	\label{Eqn:CovarSolns}
\end{equation}
which is valid because the added noise is Markovian and uncorrelated to the system. The aggregate added noise $V_{H}$ is
\begin{eqnarray*}
	V_{H} & = & \intfin{t^{\prime}}{0}{t} \intfin{t^{\prime\prime}}{0}{t} 2\Gamma\mathcal{N}\Br{t^{\prime},t^{\prime\prime}} M\Br{t-t^{\prime}} \Br{\begin{array}{cc} 0 & 0 \\ 0 & 1 \end{array}}\trans{M}\Br{t-t^{\prime\prime}} \\
	& = & \Br{\begin{array}{cc} \expect{\delta X^{2}} & \half \expect{\delta X \delta P+ \delta P \delta X} \\ \half \expect{\delta X \delta P+ \delta P \delta X} & \expect{\delta P^{2}} \end{array}},
\end{eqnarray*}
with
\begin{eqnarray*}
	\expect{\delta X^{2}} & = & \Br{2\n_{H}+1}\Sq{1+\frac{\e^{-\Gamma  t}}{\sigma^{2}}\Br{\frac{\Gamma^{2}\cos\Br{2 \sigma \omega_{M} t}-2\Gamma\sigma\omega_{M}\sin\Br{2\sigma\omega_{M} t}}{4\omega_{M}^{2}}-1}}, \\
	\expect{\delta P^{2}} & = &  \Br{2\n_{H}+1}\Sq{1+\frac{\e^{-\Gamma  t}}{\sigma^{2}} \Br{\frac{\Gamma^{2}\cos\Br{2\sigma \omega_{M} t}+2\Gamma\sigma\omega_{M}\sin\Br{2\sigma\omega_{M} t}}{4 \omega_{M}^{2}}-1}},
\end{eqnarray*}
and
\begin{equation*}
	\half \expect{\delta X \delta P+ \delta P \delta X} = \Br{2\n_{H}+1}\Sq{\frac{\Gamma \e^{-\Gamma t}}{\sigma^{2} \omega_{M}} \sin^{2}\Br{\sigma\omega_{M}  t}}.
\end{equation*}

Note that for $t \ll \omega_{M}^{-1}$ the variance of $\delta P$ goes linearly in $t$, the covariance of $\delta X$ \& $\delta P$ goes quadratically, and the variance of $\delta X$ scales cubically, \viz{}
\begin{eqnarray*}
	\expect{\delta X_{M}^{2}} & \propto & \frac{2}{3}\Gamma \omega_{M}^{2}t^{3} + \cO\Br{t^{4}}, \\
	\expect{\delta P_{M}^{2}} & \propto & 2 \Gamma t + \cO\Br{t^{2}}, \\
	\half \expect{\delta X \delta P+ \delta P \delta X} & \propto & \Gamma \omega_{M} t^{2} + \cO\Br{t^{3}}.
\end{eqnarray*}

The long-time evolution yields the steady-state covariance matrix
\[
	V_{CSS}^{\Br{\mathrm{IO}}} = \lim_{t\rightarrow \infty} V\td = \Br{2\n_{H}+1}\identity,
\]
which is easily recognizable as a thermal state with occupancy $\n_{H} \approx k_{B}T_{H}/\hbar\omega_{M}$ (for the relevant case of $k_{B}T_{H} \gg \hbar \omega_{M}$, as phonons are bosonic excitations).

\subsection*{Born--Markov master equation} \label{Sec:RWAEOMs}

The BMME is obtained by discarding non-energy-conserving terms in the independent oscillator model Hamiltonian \footnote{The bare resonance frequency must also be renormalized.}. This corresponds to making the rotating wave approximation (RWA) on the independent oscillator model \cite{SuppBowenMilburn}.

It is instructive to consider the oscillator's dynamics under the BMME. Solving \eqref{Eqn:XdotRWA} and \eqref{Eqn:PdotRWA} requires knowledge of the noise correlation functions, \viz{}
\begin{eqnarray*}
	\expect{X_{in}\td^{2}} & = & \Br{2\n_{H}+1}\delta \Br{t-t^{\prime}}, \\
	\expect{P_{in}\td^{2}} & = & \Br{2\n_{H}+1}\delta \Br{t-t^{\prime}}, \\
	\Re{\expect{X_{in}P_{in}}} & = & 0.
\end{eqnarray*}

Using these yields the evolution of the covariance matrix, which is entirely analogous to \eqref{Eqn:CovarSolns} except that
\[
	M_{H}^{\Br{\mathrm{RWA}}} = \e^{-\Gamma t/2} R\Br{\omega_{M}t}
\]
and
\[
	V_{H}^{\Br{\mathrm{RWA}}} = \Br{2\n_{H}+1}\Br{1-\e^{-\Gamma t}}\identity,
\]
where $R\Br{\omega_{M}t}$ is a rotation through an angle of $\omega_{M}t$, and $\identity$ is the identity matrix. Note that $V_{H}^{\Br{\mathrm{RWA}}}$ is proportional to the identity matrix, and at short times both diagonal elements of $V_{H}^{\Br{\mathrm{RWA}}}$ grow at first order in $t$; this is very different behavior from the independent oscillator model, where $\expect{\delta X^{2}}$ initially only increases at third order in $t$. 

Despite this, the long-time behavior in the RWA is identical to the independent oscillator model, \viz{}
\[
	V_{CSS}^{\Br{\mathrm{RWA}}} = \lim_{t\rightarrow \infty} V\td = \Br{2\n_{H}+1}\identity = V_{CSS}^{\Br{\mathrm{IO}}}.
\]

\subsection*{`Instantaneous' cold bath interaction} \label{App:NonRWABath}

Let us consider the interaction with the cold bath. As stated above, we have taken this to be instantaneously quick. In this section, we will derive the necessary relations.

To begin, consider the momentum-damped Langevin equations. We will let the damping rate into the cold bath be $\gamma$ and the thermal noise be characterized by $\n_{C}$. The amount of damping is characterized by the product $\beta = \gamma\tau_{C}$, where $\tau_{C}$ is the interaction time; thus, in order to take the `instantaneous' interaction limit, we will fix $\beta$ whilst letting $\tau_{C} \rightarrow 0$ and $\gamma \rightarrow \infty$. Large damping rates require us to use the overdamped solutions to \eqref{Eqn:Xdot} and \eqref{Eqn:Pdot}. The homogeneous part is
\[
	M_{C} = \e^{-\half\gamma \tau_{C}}\Br{\begin{array}{cc}
							\cosh\frac{\alpha \tau_{C}}{2} + \frac{\gamma}{\alpha}\sinh\frac{\alpha \tau_{C}}{2} & \frac{2\omega_{M}}{\alpha}\sinh\frac{\alpha \tau_{C}}{2} \\
							-\frac{2\omega_{M}}{\alpha}\sinh\frac{\alpha \tau_{C}}{2} & \cosh\frac{\alpha \tau_{C}}{2} - \frac{\gamma}{\alpha}\sinh\frac{\alpha \tau_{C}}{2},
						\end{array}}
\]
and the corresponding noise is
\[
	V_{C} = \Br{2\n_{C}+1}\Br{\begin{array}{cc}
		1+\frac{\e^{-\gamma \tau_{C}}}{\alpha^{2}}\Sq{ \begin{array}{c} 4\omega_{M}^{2} - \gamma^{2}\cosh\alpha \tau_{C} \\
		-\alpha \gamma \sinh \alpha \tau_{C}\end{array}} & \frac{2\gamma\omega_{M}}{\alpha^{2}}\e^{-\gamma \tau_{C}} \Sq{\cosh\alpha \tau_{C}-1} \\ \frac{2\gamma\omega_{M}}{\alpha^{2}} \e^{-\gamma \tau_{C}}\Sq{\cosh\alpha \tau_{C}-1} & 1+\frac{\e^{-\gamma \tau_{C}}}{\alpha^{2}}\Sq{\begin{array}{c}4\omega_{M}^{2} - \gamma^{2}\cosh\alpha \tau_{C}\\
		+\alpha \gamma \sinh \alpha \tau_{C}\end{array}}
	\end{array}}.
\]
In these, $\alpha = \sqrt{\gamma^{2}-4 \omega_{M}^{2}}$ is a real parameter.

Substituting $\gamma = \beta/\tau_{C}$ and taking the $\tau_{C}\rightarrow 0$ limit yields
\[
	M_{C} = \Br{\begin{array}{cc} 1 & 0 \\ 0 & \e^{-\beta}\end{array}}
\]
and
\[
	V_{C} = \Br{2\n_{C}+1}\Br{\begin{array}{cc} 0 & 0 \\ 0 & 1-\e^{-2\beta}\end{array}}.
\]
As might be expected, there is no loss on the position, so the process introduces only momentum noise.

We need to draw a correspondence with the RWA case, for a fair comparison. Performing the same process using the BMME yields
\[
	\upBr{M_{C}}{\RWA} = \e^{-\beta/2}\identity,
\]
and
\[
	\upBr{V_{C}}{\RWA} = \Br{2\n_{C}+1}\Br{1-\e^{-\beta}}\identity.
\]

Compare this to a beamsplitter; we see there is a correspondence with a beamsplitter with reflectivity $\epsilon = 1-\e^{-\beta}$. This will be our fixed number that allows us to compare the RWA and non-RWA results.

To summarise, in the RWA
\begin{eqnarray}
	\upBr{M_{C}}{\RWA} & = & \sqrt{1-\epsilon}\identity, \\
	\upBr{V_{C}}{\RWA} & = & \Br{2\n_{C}+1}\epsilon\identity,
\end{eqnarray}
and without the RWA,
\begin{eqnarray}
	M_{C} & = & \Br{\begin{array}{cc} 1 & 0 \\ 0 & 1-\epsilon \end{array}}, \\
	V_{C} & = & \Br{2\n_{C}+1}\Br{\begin{array}{cc} 0 & 0 \\ 0 & \epsilon\Br{2-\epsilon} \end{array}}.
\end{eqnarray}
In the limit of a `perfectly reflective beamsplitter' ($\epsilon = 0$), the non-RWA operation replaces the $P$ quadrature with thermal noise.

Note that the thermal occupancy $\n_{C}$ can be arbitrarily low in the RWA case without problems, whilst in the non-RWA case we must have $\n_{C} \gg 1$ to ensure that the transformation preserves the Heisenberg uncertainty relation.

\section{Approximation of cyclical steady-state occupancy} \label{App:SteadyState}

In the main text we defined the `cyclical steady-state' covariance matrix as \eqref{Eqn:SteadyState}, reproduced here for convenience.
\begin{equation}
	V_{CSS} = M_{hom}\Br{\tau}V_{CSS}\trans{M}_{hom}\Br{\tau}  + V_{add}\Br{\tau}.
	\label{Eqn:SteadyStateSup}
\end{equation}
We can cast this in the typical form of a Sylvester equation by noting that $M_{hom}^{-1}$ does exist for any non-negative, finite delay time $\tau$. Thus
\begin{equation}
	M_{hom}^{-1}V_{CSS} - V_{CSS}\trans{M}_{hom} = M_{hom}^{-1}V_{add}.
	\label{Eqn:Sylvester}
\end{equation}
This can be solved with standard computational packages, or indeed analytically. The general solution to \eqref{Eqn:Sylvester} is \cite{SuppMatrixCookbook}
\[
	\mathrm{vec}\Cu{V_{CSS}} = \Br{\identity\otimes \Br{M_{hom}^{-1}}-M_{hom}\otimes \identity}^{-1}\mathrm{vec}\Cu{M_{hom}^{-1}V_{add}},
\]
where $\otimes$ is the Kronecker product and $\mathrm{vec}\Cu{\cdots}$ is the vectorisation operation.

\begin{figure}
	\centering
	\def\svgwidth{0.6\columnwidth}
	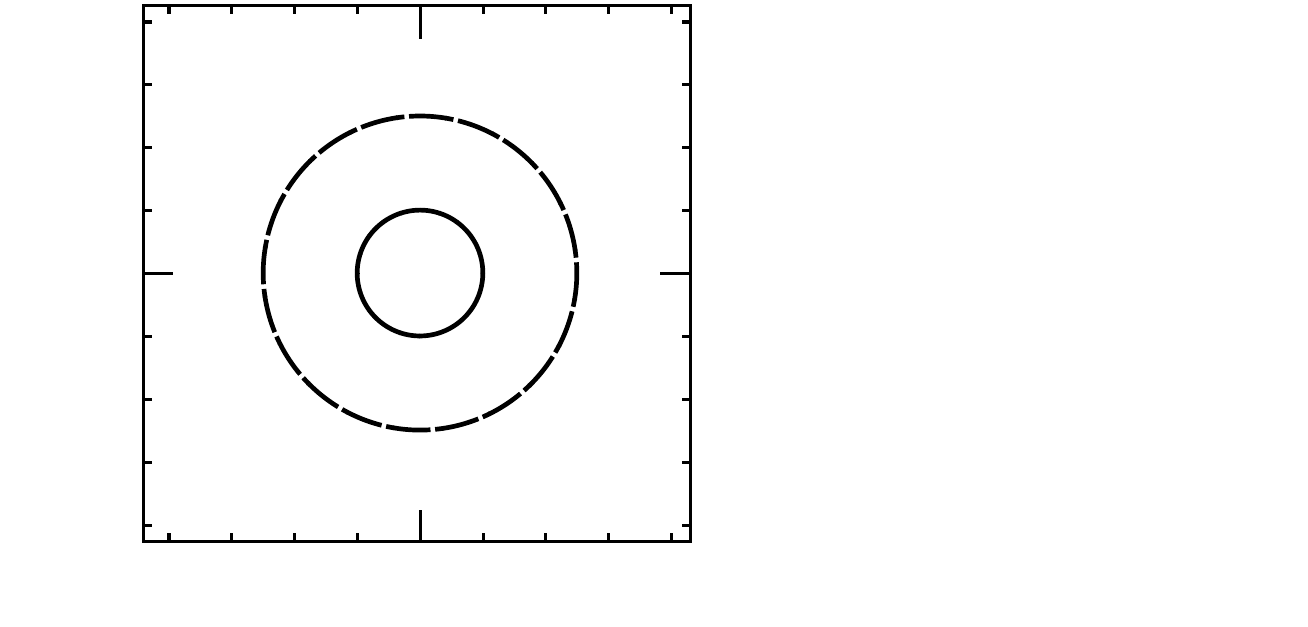
	\caption{\label{Fig:Ellipses}
	Comparison of steady-state Wigner function contours calculated using momentum-dependent damping (solid lines) and in the RWA (dashed). The system parameters are $\omega_{M} = 10^{6}$~Hz, $Q = 10^{6}$, $\n_{H} = 4\times 10^{4}$, $\n_{C} = 10^{2}$, and $\epsilon = \pi \times 10^{-9}$, with the squeezing parameters being $\sqrt{2}$ and $2$ for \textit{a}) and \textit{b}) respectively. Small squeezing strengths have been chosen because the relative size of $V_{CSS}^{\Br{\mathrm{RWA}}}$ grows rapidly with $\mu$.
	}
\end{figure}

As seen in Supp.~Fig.~\ref{Fig:Ellipses}, the steady-state Wigner function is essentially symmetrical between position and momentum. Therefore, the state is thermal and characterized by an effective occupancy
\[
	\n_{CSS} = \frac{\sqrt{\det\Cu{V_{CSS}}}-1}{2}.
\]

The occupancy $\n_{CSS}$ can in principle be calculated entirely analytically, but the resulting expression is immensely unwieldy. Instead, we will find an approximate expression which is accurate in the region of interest ($\Cu{Q,\n_{H}} \gg 1$, $\Cu{\epsilon, \epsilon \n_{C}, \omega_{M} \tau} \ll 1$).

For simplicity, let us first consider the perfect squeezing ($\epsilon = 0$) case. Each matrix may be calculated analytically. We then expand $\mathrm{vec}\Cu{V_{CSS}}$ to second order in $\tau$, calculate the determinant, and truncate the result at $\cO\Br{\tau^{2}}$ and $\cO\Br{\Gamma}$. Discarding small terms (that are not boosted by $\mu$) then gives
\begin{equation}
	\n_{CSS}|\sb{\epsilon=0} \approx \bar{n}_{H} \Br{\frac{1}{\mu^{2}} + \frac{4\pi^{2}\omega_{M}^{2}}{3\omega_{ap}^{2}} \mu^{2}}.
	\label{Eqn:nSSNoCold}
\end{equation}

The coefficient of $\mu^{2}$ in \eqref{Eqn:nSSNoCold} can in fact be related directly to properties of the added noise. To see this, let us ask which value of $\mu$ minimizes the energy contained by the added noise $V_{add}$. In this case ($\epsilon = 0$) we have $V_{add} = S_{2}V_{H}S_{2}^{\mathrm{T}}$; thus we wish to select the value of $\mu$ which minimizes $\Tr{S_{2}^{\mathrm{T}}S_{2} V_{H}}$. Calculating this yields
\[
	\mu_{opt}^{4}|\sb{\epsilon = 0} = 3\Br{\frac{\omega_{ap}}{2\pi\omega_{M}}}^{2},
\]
which is the inverse of the coefficient of $\mu^{2}$ in \eqref{Eqn:nSSNoCold}.

Similarly, consider the $\mu = 1$ value of $\n_{CSS}|\sb{\epsilon = 0}$, which is approximately $\n_{H}$.

If we rewrite $\n_{CSS}$ as
\[
	\n_{CSS}|\sb{\epsilon = 0} = \Br{\n_{CSS}|\sb{\mu = 1}} \Br{\frac{1}{\mu^{2}} + \frac{\mu^{2}}{\mu_{opt}^{4}|\sb{\epsilon = 0}}}
\]
then---guided by our numerical calculations---we can make the educated guess that the same form holds when the squeezers become imperfect.

The first factor $\n_{CSS}|\sb{\mu = 1}$ needs to be generalized to there being two baths. The most likely estimate is the typical equilibrium occupancy expected from detailed balance, \viz{}
\[
	\n_{CSS}|\sb{\mu = 1} \rightarrow \frac{\Gamma \n_{H} + \gamma_{eff} \n_{C}}{\Gamma + \gamma_{eff}},
\]
The effective coupling rate to the cold bath is $\gamma_{eff}$, which is taken to be $\gamma_{eff} \approx 2\epsilon/\tau = \epsilon \omega_{ap}/\pi$ because the loss of $2\epsilon$ occurs over every timestep $\tau$.

Secondly, we replace $\mu_{opt}|\sb{\epsilon = 0}$ with the value of $\mu$ that minimizes the trace of $V_{add}$ \textit{including} noise from the cold bath. This may be analytically found to be
\[
	\mu_{opt} = \frac{1 -2\epsilon\cos 2\omega_{M}\tau}{{2\omega_{M}\tau-\sin 2\omega_{M}\tau}} \Sq{
	2\omega_{M}\tau + \sin 2\omega_{M}\tau + 4\pi\frac{\omega_{M}}{\omega_{ap}}\frac{\gamma}{\Gamma}\frac{2\n_{C}+1}{2\n_{H}+1}\Br{1-\frac{\Gamma}{2\omega_{M}}\sin 2\omega_{M}\tau}
	},
\]
which in turn can be reduced to the expression appearing in the main text (\eqref{Eqn:MuOpt}) if $\n_{C} \gg 1$, and $\tau \rightarrow 0$.

By this route we arrive at \eqref{Eqn:SteadyOccupancy} of the main text. As seen in Fig.~\ref{Fig:SurfaceOccupancy}~\textit{c}), this is a reasonable approximation in the parameter regime of interest.

Note that none of the bounds of thermodynamic cycles (Appendix~\ref{App:ThermoCycles}) rely on this particular form of $\n_{CSS}$; they simply require a symmetrical steady-state Wigner function.

\section{Comparison to dedicated cooling procedures} \label{Sec:OtherCooling}

One interesting achievement of our protocol is the ability to cool the oscillator to below the cold bath temperature during parts of the cycle. In this section, we compare this behavior to key optomechanical cooling protocols--- feedback cooling (`cold damping') \cite{SuppPoggio2007} and resolved-sideband cooling---with comparable resources. As we will see, both conventional schemes are unable to cool the mechanical oscillator to below the cold bath temperature (set by the purity of the optical field); however, they do achieve comparable cooling with far fewer photons.

\subsection*{Feedback cooling} \label{Sec:FeedbackCooling}

Feedback cooling uses an optical readout of the mechanical position to apply a friction force to the oscillator. In the case of continuous-wave probe light with an occupancy of $\n_{C}$ and cooperativity of $C$, the oscillator's position variance obeys
\[
	V_{XX,FB} = \frac{2}{1+G}\Br{\n_{H}+\half + \Br{2\n_{C}+1}\Br{C+\frac{G^{2}}{16 C}}},
\]
where $G$ is the feedback gain that quantifies the strength of feedback. This expression reproduces existing results in the case of coherent light ($\n_{C} = 0$) \cite{SuppPinard2001,SuppPoggio2007,SuppBennett2014}.

Minimising the variance with respect to $G$ gives
\begin{eqnarray*}
		V_{XX,FB}^{\mathrm{min}} & = & \frac{2\n_{C}+1}{4C} \Br{\sqrt{\begin{array}{c}\Br{2\n_{C}+1}\Br{1+16 C} \\ {}+8 C \Br{2\n_{H}+1} \end{array}} - 1}.
\end{eqnarray*}
Importantly, even in the high-cooperativity limit, this cannot be reduced below the cold bath temperature;
\[
	V_{XX,FB}^{\mathrm{min}} > \Br{2\n_{C}+1},
\]
\textit{ergo}, feedback cooling cannot make the oscillator colder than the temperature set by the purity of the optical probe.

To choose a fair value of $C$ for comparison with the cooling achieved in our scheme, we set the CW pump power used in the feedback scheme to be equal to the average input power of the squeeze--rotate--unsqueeze protocol, \viz{}
\[
	C = \frac{\omega_{ap}}{\omega_{M} }\frac{8Q}{\pi} \Br{\frac{g_{0}}{\kappa}}^{2} \sum\sb{j} N\sb{j},
\]
where $j$ runs over all ten pulses involved in a single iteration of the protocol. Using the parameters in Table~\ref{Tab:Params}, $C$ is large enough to ground-state cool the oscillator even from room temperature ($8C > \n_{H}$) if the probe light is coherent.

\subsection*{Sideband cooling} \label{Sec:ResolvedCooling}

Sideband cooling is a well-established means of leveraging dynamical back-action to remove phonons from an optomechanical oscillator \cite{SuppSchliesser2008,SuppSchliesser2009,SuppChan2011a,SuppTeufel2011}.

In the traditional implementation with coherent light, efficient sideband cooling requires that the cavity bias the optomechanical interaction towards anti-Stokes scattering. This is most effective with a detuned drive and a sideband-resolved cavity \ie{} $\kappa < \omega_{M}$ \cite{SuppSchliesser2008}. The oscillator then reaches a minimum effective temperature of $\n_{eff,SB} \approx \Br{\kappa/4\omega_{M}}^{2} >0$.

Recent work \cite{SuppClark2017} has relaxed the $\kappa < \omega_{M}$ requirement by using squeezed pump light. In fact, the optomechanical Stokes scattering in such a system can be coherently cancelled using a finite amount of input squeezing. This is caused by the optomechanical Kerr nonlinearity. Near-ground-state cooling has been achieved using this scheme \cite{SuppClark2017}.

From Eq.~(6) of \cite{SuppClark2017}, we find that, with the optimum amount of squeezing and drive detuning, the oscillator is cooled to a minimum variance of
\[
	V_{XX,SB}^{\mathrm{min}} = 2\n_{C}\Br{1+\frac{\kappa^{2}}{2\omega_{M}^{2}}}+1.
\]
Again, we see that the thermal occupancy (purity) of the probe light sets a hard lower limit to the mechanical occupancy, unlike our proposed scheme.

\section{Bounds on thermodynamic cycles} \label{App:ThermoCycles}

Consider a complete cycle of the squeeze--rotate--unsqueeze protocol with lossy squeezing as described in the main text. Let us divide up the protocol into the following steps (\cf{} Fig.~1 of main text):
\begin{eqnarray*}
	V^{\Br{1}} & = & S_{1} V_{CSS} \trans{S_{1}}, \\
	V^{\Br{2}} & = & M_{C} V^{\Br{1}} \trans{M_{C}} + V_{C}, \\
	V^{\Br{3}} & = & M_{H}V^{\Br{2}}\trans{M}_{H}, \\
	V^{\Br{4}} & = & S_{2}V^{\Br{3}}\trans{S_{2}}, \\
	V_{CSS} & = & M_{C} V^{\Br{4}} \trans{M_{C}} + V_{C} .
\end{eqnarray*}

The squeezing steps perform work, allowing us to identify the total work (in quanta) as
\[
	W = \frac{1}{4}\Tr{V^{\Br{4}}-V^{\Br{3}} + V^{\Br{1}}-V_{CSS}}.
\]

Similarly,
\begin{eqnarray*}
	Q_{H} & = & \frac{1}{4}\Tr{V^{\Br{3}}-V^{\Br{2}}}, \\
	Q_{C} & = & \frac{1}{4}\Tr{V_{CSS}-V^{\Br{4}}+V^{\Br{2}}-V^{\Br{1}}}.
\end{eqnarray*}

\subsection*{Criterion for heat engine regime} \label{App:EnginePhase}

The system behaves as a heat engine if it receives heat from the hot bath ($Q_{H}>0$) and produces output work ($W < 0$). Any excess entropy is dumped into the cold bath.

Our numerical calculations show that the engine regime overlaps very well with the parameter regime where the ratio of eigenvalues of $V^{\Br{3}}$ is larger than that of $V^{\Br{2}}$, \ie{} where the asymmetry of the lossy evolution step is strong enough to actually \textit{increase} the eccentricity of contours of the Wigner function over a short timestep.

For a vacuum state operated on by $S\sb{1}\Br{\mu}$ we have the ratio of eigenvalues $\kappa = \expect{P^{2}}/\expect{X^{2}} = \mu^{4}$. It is easily seen that $\kappa$ is in fact a function of
\[
	K =\frac{\Tr{V}}{\sqrt{\mod{V}}},
\]
which is directly proportional to the product of the energy ($\propto \Tr{V}$) and purity $\mod{V}^{-1/2}$ of the state. Thus we can use $K$ as a proxy for the ratio of eigenvalues.

To simplify the calculation of $K^{\Br{2}}$ and $K^{\Br{3}}$ we will consider the high-$Q$, low-$\epsilon$ limit and set $\sigma = \sqrt{1-\Gamma^{2}/4\omega_{M}^{2}} = 1$. We expand the resulting inequality $\upBr{K}{3} > \upBr{K}{2}$ around small $t$ to first order, yielding
\[
	\Br{\begin{array}{c}
	\n_{CSS}\mu^{2}\Br{\mu^{2}\n_{CSS}-\n_{H}}\Br{\mu^{8}-1}+\\
	2\epsilon\Br{\n_{C}-2\n_{CSS}\mu^{2}}\Br{2\n_{CSS}\mu^{6}+\n_{H}\Br{\mu^{8}-2\mu^{4}-1}}\end{array}} < 0
\]
which in most cases of interest ($\Cu{\epsilon,\epsilon\n_{C}} < 1 \ll \n_{H}$) can be replaced by the simplified condition
\[
	\Br{\mu^{2}\n_{CSS}-\n_{H}}\Br{\mu^{8}-1} <0.
\]
Thus
\begin{eqnarray*}
	\n_{H} > \mu^{2}\n_{CSS}, & \;\;\; & \mu >1, \\
	\n_{H} < \mu^{2}\n_{CSS}, & \;\;\; & \mu <1.
\end{eqnarray*}

To interpret these results it suffices to consider an extremely simple model. Over a very short timestep the position noise is essentially unchanged by damping, whilst the momentum is attenuated and heated. Thus for $\mu > 1$ ($\mu < 1$) the ratio of eigenvalues increases if $V_{PP}$ increases (decreases). The momentum noise at the beginning of the step is approximately $\upBr{V_{PP}}{2} \approx 2\n_{CSS}\mu^{2}\Br{1-\epsilon}^{2}$, and at the end $\upBr{V_{PP}}{3} \approx \Br{1-\Gamma t}\upBr{V_{PP}}{2} + 2\Gamma t \n_{H}$. Thus, $\Delta V_{PP} \approx 2\Gamma t \n_{H}-2\Gamma t\n_{CSS}\Br{1-\epsilon}^{2}$. Requiring that $\Delta V_{PP}$ be positive (negative) yields the $\mu > 1$ ($\mu < 1$) results given above.

\subsection*{Criterion for refrigeration regime} \label{App:RefrigerationPhase}

In the refrigeration regime, the resonator uses input work ($W > 0$) to extract heat from the cold reservoir ($Q_{C} > 0$) and push it into the hot reservoir ($Q_{H} < 0$).

Consider $Q_{C} > 0$. Thus
\[
	\Tr{V_{CSS}-V^{\Br{4}}+V^{\Br{2}}-V^{\Br{1}}} > 0.
\]
This can be written entirely in terms of $V_{CSS}$ and $V_{C}$, \viz{}
\[
		\Tr{\Sq{\identity - \trans{M_{C}^{-1}}M_{C}^{-1} + \trans{S_{1}}\Br{\trans{M_{C}}M_{C}-\identity}S_{1}} V_{CSS}}
	+\Tr{\Sq{\identity + \trans{M_{C}^{-1}}M_{C}^{-1}}V_{C}} > 0.
\]

Using the fact that $V_{CSS} \approx \Br{2\n_{CSS}+1}\identity$, we obtain
\[
	\n_{C}\Br{2-2\epsilon+\epsilon^{2}} > \n_{CSS}\Br{1+\Br{1-\epsilon}^{2}\mu^{2}}.
\]
In the limit that $\epsilon \ll 1$ this becomes equal to \eqref{Eqn:FridgeCondn} given in the main text.

\section{No-go theorems in the RWA} \label{App:NoGo}

The following no-go theorems are based on the complete analytical result for the steady-state covariance matrix in the RWA.

\subsection*{No-go theorem for heat engines in the RWA} \label{App:NoGoEngine}

We calculated the exact criterion for heat engine function in the RWA using Wolfram Mathematica. The expression $W^{\Br{\mathrm{RWA}}} < 0$  factorises into a form involving three factors, two of which are always positive. The final factor is only negative when
\begin{equation}
	\mu^{4} + B \mu^{2} + 1 < 0,
	\label{Eqn:WorkCondition}
\end{equation}
where
\[
	B = \frac{a \Br{2\n_{H}+1} + b\Br{2\n_{C}+1}}{c \Br{2\n_{H}+1} + d\Br{2\n_{C}+1}},
\]
with $a$, $b$, $c$, and $d$ being functions of $\epsilon$, $\tau$, $\Gamma$, and $\omega_{M}$, \viz{}
\begin{eqnarray*}
	a & = & \Br{\e^{\Gamma \tau}-1} \Br{\e^{\Gamma \tau} - \Br{1-\epsilon}^2} \times \\
	& & \Br{\e^{\Gamma \tau} + \Br{1-\epsilon}^3 - \Br{1 - \epsilon}\Br{1 + \e^{\Gamma \tau} - \epsilon} \cos 2\omega_{M} \tau} \csc^{2} \omega_{M}\tau, \\
	b & = & \left[\begin{array}{c}\e^{3 \Gamma \tau} - 2 \Br{1 - \epsilon}^5 + \e^{2 \Gamma \tau} \epsilon + \e^{\Gamma \tau} \Br{1-\epsilon}^2 \Br{1 - \epsilon\Br{3-\epsilon}} - \\
	\Br{1 - \epsilon} \Br{\e^{2 \Gamma \tau} \Br{3 - 2 \epsilon} - \Br{1 - \epsilon}^3 + \e^{\Gamma \tau} \Br{\epsilon\Br{5 - 3 \epsilon}-2}) \cos 2\omega_{M}\tau}\end{array}\right] \\
	& & \times\epsilon \csc^{2}\omega_{M} \tau, \\
	c & = & \Br{\e^{\Gamma \tau}-1}\Br{1-\epsilon} \Br{\e^{\Gamma \tau} + \Br{1-\epsilon}^2} \Br{\e^{\Gamma \tau}+\epsilon-1},\\
	d & = & \epsilon\Br{1 - \epsilon}\Br{\e^{\Gamma \tau} + \Br{1 - \epsilon}^2}  \Br{\e^{\Gamma \tau} + \epsilon - 1}.
\end{eqnarray*}
Clearly $c$ and $d$ are non-negative. Let us now consider the signs of $a$ and $b$.

The factor $a$ is positive if
\begin{equation}
	\e^{\Gamma \tau} + \Br{1-\epsilon}^3 > \Br{1 - \epsilon}\Br{1 + \e^{\Gamma \tau} - \epsilon} \cos 2\omega_{M} \tau.
	\label{Eqn:aFactor}
\end{equation}
The right hand side of this inequality is never larger than $\Br{1 - \epsilon}\Br{1 + \e^{\Gamma \tau} - \epsilon}$ (\ie{} set $\cos 2\omega_{M} \tau = 1$); thus if we can satisfy $\e^{\Gamma \tau} + \Br{1-\epsilon}^3 > \Br{1 - \epsilon}\Br{1 + \e^{\Gamma \tau} - \epsilon}$ we can always satisfy Supp.~\eqref{Eqn:aFactor}. Rearranging $\e^{\Gamma \tau} + \Br{1-\epsilon}^3 > \Br{1 - \epsilon}\Br{1 + \e^{\Gamma \tau} - \epsilon}$ yields
\[
	\e^{\Gamma \tau} > \Br{1-\epsilon}^{2},
\]
which is always true. This means that Supp.~\eqref{Eqn:aFactor} is always satisfied, and $a$ is always positive.

The remaining coefficient $b$ is positive when the term in square brackets is positive, \viz{}
\begin{equation}
	\Sq{\begin{array}{c} \e^{3 \Gamma \tau} - 2 \Br{1 - \epsilon}^{5} + \epsilon\e^{2 \Gamma \tau} + 
  \e^{\Gamma \tau} \Br{1 - \epsilon}^{2} \Br{1 -\epsilon\Br{3 - \epsilon}} - \\
	\Br{1 - \epsilon} \Cu{\e^{2 \Gamma \tau} \Br{3 - 2 \epsilon} - \Br{1 - \epsilon}^{3} + \e^{\Gamma \tau} \Br{\epsilon\Br{5 - 3 \epsilon}-2}} \cos 2 \omega_{M} \tau \end{array}} > 0.
	\label{Eqn:bFactor}
\end{equation}
We begin by noting that the term in braces (in Supp.~\eqref{Eqn:bFactor}) is always positive. This may be proven as follows:

\begin{itemize}
\item Convert the term ($\e^{2 \Gamma \tau} \Br{3 - 2 \epsilon} - \Br{1 - \epsilon}^{3} + \e^{\Gamma \tau} \Br{\epsilon\Br{5 - 3 \epsilon}-2}$) into a quadratic in $\lambda = \e^{\Gamma \tau}$.
\item Solve for the roots of the quadratic. The largest root is
\[
	\lambda_{R} = \frac{1-\epsilon}{6-4\epsilon}\Sq{2-3\epsilon-\sqrt{16-32 \epsilon+17\epsilon^{2}}}.
\]
\item Consider the inequality $\lambda_{R} > 1$. It simplifies to $\epsilon^{2}-6\epsilon + 6 < 0$.
\item Calculate the roots of this quadratic, which are $\epsilon = 3\pm \sqrt{3}$, both of which are larger than $1$.
\item Conclude that the quadratic condition on $\epsilon$ is never satisfied for physical values of $\epsilon$; thereby see that $\lambda_{R}<1$ $\forall$ $\epsilon \in \Br{0,1}$.
\item Use the definition of $\lambda$ to see that $\lambda>1$ for all $\Gamma \tau$ of interest, and thus this $\lambda_{R}$ is not physical.
\item Conclude that the term in Supp.~\eqref{Eqn:bFactor} is always positive.
\end{itemize}

With this we can see that
\[
	\begin{array}{c} \Br{1 - \epsilon} \Cu{\e^{2 \Gamma \tau} \Br{3 - 2 \epsilon} - \Br{1 - \epsilon}^{3} + \e^{\Gamma \tau} \Br{\epsilon\Br{5 - 3 \epsilon}-2}} > \\
	\Br{1 - \epsilon} \Cu{\e^{2 \Gamma \tau} \Br{3 - 2 \epsilon} - \Br{1 - \epsilon}^{3} + \e^{\Gamma \tau} \Br{\epsilon\Br{5 - 3 \epsilon}-2}} \cos 2 \omega_{M} \tau \end{array}.
\]
which prompts us to consider the simpler inequality
\begin{equation}
	\begin{array}{c} \e^{3 \Gamma \tau} - 2 \Br{1 - \epsilon}^{5} + \epsilon\e^{2 \Gamma \tau} + 
  \e^{\Gamma \tau} \Br{1 - \epsilon}^{2} \Br{1 -\epsilon\Br{3 - \epsilon}} > \\
	\Br{1 - \epsilon} \Cu{\e^{2 \Gamma \tau} \Br{3 - 2 \epsilon} - \Br{1 - \epsilon}^{3} + \e^{\Gamma \tau} \Br{\epsilon\Br{5 - 3 \epsilon}-2}} \end{array},
	\label{Eqn:bFactor2}
\end{equation}
equivalent to considering the right hand side with $\cos 2 \omega_{M}\tau = 1$. This new condition factorises rather straightforwardly, yielding
\[
	\Br{\e^{\Gamma \tau}-\Br{1-\epsilon}^{2}}^{2}\Br{\e^{\Gamma \tau}+2\epsilon -1} > 0,
\]
which is obviously satisfied. Thus we conclude that $b>0$ always.

Since we have established that $a$, $b$, $c$, and $d$ are all positive we know $B$ is positive. Consider then Supp.~\eqref{Eqn:WorkCondition}, which is quadratic in $\mu^{2}$. There are only solutions to Supp.~\eqref{Eqn:WorkCondition} when the determinant of the left hand side, $B^{2}-4$, is non-negative. This gives $B\geq 2$. Rearranging this yields the condition
\[
	\e^{2 \Gamma \tau} + (1 - \epsilon)^4 - 2 \e^{\Gamma \tau} \Br{1 - \epsilon}^{2} \cos 2\omega_{M}\tau > 0,
\]
which is seen to be quadratic in $\lambda = \e^{\Gamma \tau}$. The determinant of the left hand side is negative, thus there are no real roots. This means that this conditions is always satisfied, and $B$ is always at least $2$.

We may thus calculate that $\mu^{4} + B \mu^{2} + 1 < 0$ is satisfied when
\begin{equation}
	\frac{-1}{2}\Br{\sqrt{B^{2}-4}+B} < \mu^{2} < \frac{1}{2}\Br{\sqrt{B^{2}-4}-B}.
	\label{Eqn:Condition}
\end{equation}
This clearly has no solution for real $\mu$ because both the upper and lower limits of Supp.~\eqref{Eqn:Condition} are negative. This implies $W^{\Br{\mathrm{RWA}}} < 0$ cannot be satisfied, and hence that there is no heat engine regime in the RWA. Note that we have not taken any limits during this derivation, nor have we approximated $V_{CSS}$ as diagonal, so our no-go theorem holds for arbitrary $\epsilon \in \Br{0,1}$ and any  positive $\Gamma < \omega_{M}/2$.

\subsection*{No-go theorem for refrigeration in the RWA} \label{App:NoGoFridge}

Consider the criterion for refrigeration, namely $\Tr{V_{CSS}-V_{4}+V_{2}-V_{1}} > 0$. Substituting the appropriate covariance matrices and using the cyclical property of the trace gives
\[
	2\frac{2-\epsilon}{1-\epsilon}\Br{2\n_{C}+1} > \Tr{\Br{ \begin{array}{cc} \mu^{-2}+\frac{1}{1-\epsilon} & 0 \\ 0 & \mu^{2}+\frac{1}{1-\epsilon} \end{array} }V_{CSS}}.
\]
This can be re-written
\begin{equation}
	\frac{2\n_{C}+1}{2\n_{H}+1} \alpha < \beta.
	\label{Eqn:FridgeCondnRWA}
\end{equation}
The coefficients $\alpha$ and $\beta$ are relatively high-order polynomials in $\epsilon$ and $\lambda$, so we will consider the limit of small $\omega_{M}t$ and $\epsilon$. Then
\begin{eqnarray*}
	\alpha & = & 4 \epsilon^{5} \mu^{4} \Br{2-\epsilon}^{4} \Sq{2 \epsilon \Br{\mu^2-1}^2 + \Gamma \tau \Br{5 - 14 \mu^2 + 5 \mu^4 + \epsilon \Br{1 + \mu^2}^2}} \\
	\beta & = & 2 \epsilon^{5} \mu^{4} \Br{2-\epsilon}^{4} \Gamma \tau \Sq{\epsilon \Br{1 + 6 \mu^2 + \mu^4}-2 \Br{1 + \mu^2}^2}.
\end{eqnarray*}
The sign of each coefficient is controlled by the terms in square brackets. $\alpha$ does have real zeros, but these only occur for negative values of $\mu^{2}$, which are unphysical. Thus $\alpha > 0$ for all $\mu > 1$. The (square) bracketed term in $\beta$ is quadratic in $\mu^{2}$. Calculating its determinant makes it clear that $\beta < 0$ for all physical $\mu$. This means that the left hand side of Supp.~\eqref{Eqn:FridgeCondnRWA} is always positive, whilst the right is always negative, thereby ensuring that the condition cannot be met. We conclude that there is no refrigeration in the RWA (at least in the small $\epsilon$, high-$Q$ regime). Our numerical calculations support this conclusion.

\section{Rotated pulsed squeezer} \label{App:SixPulse}

The rapid optomechanical squeezer proposed in \cite{SuppBennett2018} is capable of position (phase space angle of $0$) or momentum ($\pi/2$) squeezing, but not intermediate angles. The latter is an important component of our squeeze--rotate--unsqueeze protocol. 

To create a rotated squeezer, let us begin by considering the QND sequence
\[
	M_{tot,rot} = M\Br{\chi_{3}^{\prime},\theta,0}M\Br{\chi_{2}^{\prime},\pi/2+\theta,\pi/2}M\Br{\chi_{1}^{\prime},\theta,0},
\]
where the Hamiltonian that generates $M\Br{\chi^{\prime},\theta_{M},\theta_{L}}$ is of the form
\[
	H \sim \chi^{\prime} X_{M}\Br{\theta_{M}}X_{L}\Br{\theta_{L}}.
\]
If we set the interaction strengths according to $\chi_{3}^{\prime} = -\chi_{1}^{\prime}/\mu$, and $\chi_{2}^{\prime} = \Br{1-\mu}/\chi_{1}^{\prime}$ we achieve the transformation
\begin{eqnarray*}
	X_{M} & \rightarrow & \Sq{\frac{1}{2\mu} + \frac{\mu}{2} + \Br{\frac{\mu}{2}-\frac{1}{2\mu}}\cos 2\theta}X_{M} + \Br{\mu - \frac{1}{\mu}}\cos\theta\sin\theta P_{M} + \frac{\mu-1}{\chi_{1}^{\prime}}\cos\theta P_{L}, \\
	P_{M} & \rightarrow & \Sq{\frac{1}{2\mu} + \frac{\mu}{2} - \Br{\frac{\mu}{2}-\frac{1}{2\mu}}\cos 2\theta}P_{M} + \Br{\mu - \frac{1}{\mu}}\cos\theta\sin\theta X_{M} +\frac{\mu-1}{\chi_{1}^{\prime}}\sin\theta P_{L}.
\end{eqnarray*}
Up to the optical contributions---which can be reduced by squeezing the input phase fluctuations $P_{L}$---this is a rotated squeezer, as desired.

\begin{figure}
	\centering
	\def\svgwidth{0.7\columnwidth}
	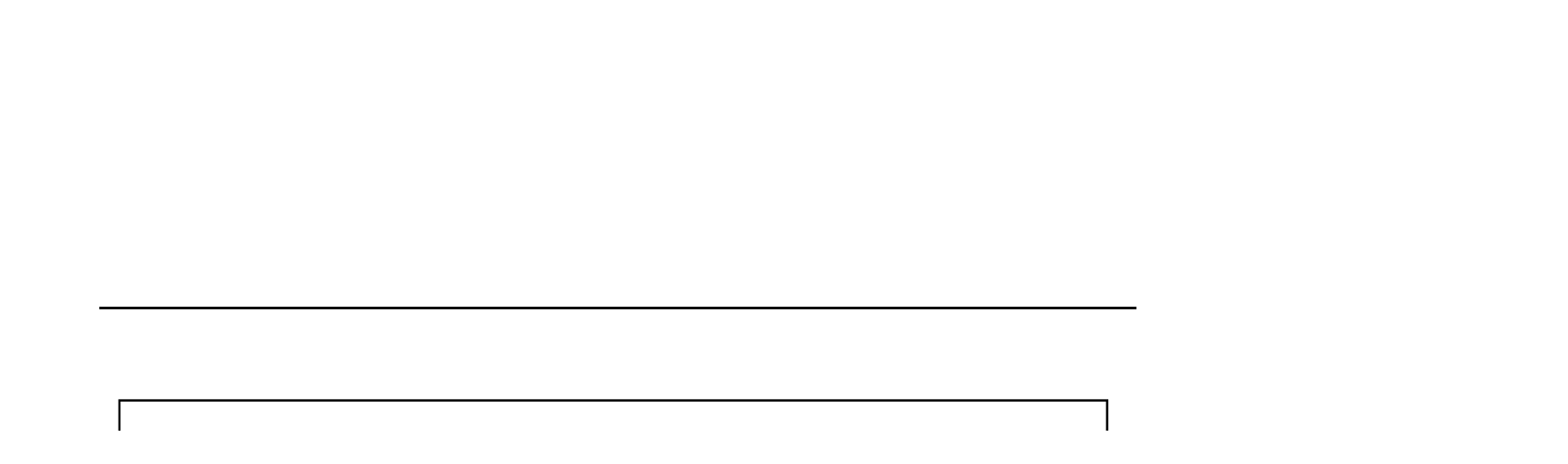
	\caption{\label{Fig:SixPulser} Approximate rotated squeezer $S_{2}$ based on three QND interactions, similar to the protocol in \cite{SuppBennett2018}. The top row shows the desired sequence; the lower shows the sequence approximated---up to an irrelevant optical rotation---using pulsed optomechanical interactions (purple), mechanical rotations (red), and optical rotations (teal). The initial state has $\mu = 3$ and $\varphi = \pi/20$. Numerically optimized parameters are given in Table~\ref{Tab:NumParams}.}
\end{figure}

Such a sequence cannot be directly executed using pulsed optomechanics, which gives access to only $X_{M}X_{L}$ interactions. To remedy this, we make use of \cite{SuppKhosla2017} to approximate each interaction using two optomechanical pulses separated by a short mechanical delay. The resulting sequence is shown in Fig.~\ref{Fig:SixPulser} (lower). The pulse strengths and rotation angles were then numerically optimized using an iterative grid search algorithm, subject to the restrictions:
\begin{itemize}
	\item the optical squeezing satisifes $0.1 < V_{L} < 1$;
	\item $\mod{\chi\sb{j}} > 0.1 \; \forall \; j\in\Cu{1,3,5}$ (no vanishing pulses);
	\item pulses 2, 4, and 6 were constrained according to \cite{SuppKhosla2017}, \ie{}
		\begin{eqnarray*}
			\chi_{2} & = & \frac{\chi_{1}\tan\theta}{\tan\theta \cos\theta_{1}-\sin\theta_{1}}, \\
			\chi_{4} & = & \frac{\chi_{3}\cot\theta}{\cot\theta \cos\theta_{2}+\sin\theta_{2}}, \\
			\chi_{6} & = & \frac{\chi_{5}\tan\theta}{\tan\theta \cos\theta_{3}-\sin\theta_{3}};
		\end{eqnarray*}
	\item $\theta\sb{k} > \theta/100 \; \forall \; k\in\Cu{1,2,3}$ (no vanishing rotations); and
	\item $\theta_{1}+\theta_{2}+\theta_{3} \leq \theta = \varphi$ (executed over a timescale not exceeding the damped evolution time).
\end{itemize}

As a representative result, consider the case of transforming a squeezed vacuum state ($\mu = 3$, $\varphi = \pi/20$: left panel of Fig.~\ref{Fig:SixPulser}) back to the vacuum. A numerically-optimized pulse sequence that achieves this is given in Table~\ref{Tab:NumParams}. The fidelity of the output state (right panel of Fig.~\ref{Fig:SixPulser}) is $\mathcal{F} = 2/\sqrt{\mod{V_{out}+\identity}} \approx 98\%$.

\begin{table}
\centering
	\begin{tabular}{ll|ll}
		$\chi_{1}$ & $0.2531$ & $\theta_{1}/\pi$ & $0.0480$ \\
		$\chi_{2}$ & $6.2052$ & $\theta_{2}/\pi$ & $3.1831\times 10^{-5}$\\
		$\chi_{3}$ & $0.1000$ & $\theta_{3}/\pi$ & $3.1831\times 10^{-5}$\\
		$\chi_{4}$ & $0.1000$ & $\theta_{opt}/\pi$ & $1.9537$ \\
		$\chi_{5}$ & $3.8516$ & $V_{L}$ & $0.1$ \\
		$\chi_{6}$ & $3.8540$ & $\mathcal{F}$ & $0.9897$ 
	\end{tabular}
	\caption{\label{Tab:NumParams} Numerically optimized parameters for a rapid rotated squeezer ($S_{2}$) in the absence of decoherence. The input was a pure squeezed state with $\mu = 3$ and $\varphi = \pi/20$. Further details are given in text.}
\end{table}

The extended rapid optomechanical squeezer presented here is rather rudimentary and could likely be improved by more carefully considering the optical rotation angles used in the protocol (\eg{} to mitigate the unwanted Kerr effect that arises over multiple pulsed interactions by including variable optical rotation angles \cite{SuppBennett2018}). This is outside the scope of this work. However, even this simple model demonstrates that the rapid, high-fidelity, rotated squeezer is possible in principle.

\end{document}